\documentclass[10pt,preprint]{emulateapj}

\usepackage{CJK}
\usepackage[flushleft]{threeparttable}
\usepackage{multirow}

\usepackage{amsmath}

\shorttitle{Nebular spectra of type IIb supernovae}
\shortauthors{Fang $\&$ Maeda}

\begin{document}
\begin{CJK*}{UTF8}{gbsn}

\title{The origin of the H$\alpha$-like structure in nebular spectra of type IIb supernovae}
\author{Qiliang Fang (方其亮)\altaffilmark{1,2,3}, Keiichi Maeda\altaffilmark{1}}
\altaffiltext{1}{Department of Astronomy, Kyoto University, Kitashirakawa-Oiwake-cho, Sakyo-ku, Kyoto 606-8502, Japan}
\altaffiltext{2}{Kavli Institute for Astronomy and Astrophysics, Peking University, Yi He Yuan Lu 5, Hai Dian District, Beijing 100871, China}
\altaffiltext{3}{Department of Astronomy, School of Physics, Peking University, Yi He Yuan Lu 5, Hai Dian District, Beijing 100871, China}

\begin{abstract}
We investigate the origin of the H$\alpha$-like structure seen in late-phase nebular spectra of type IIb supernovae (SNe IIb) at $\sim 200$ days after the explosion. We compare the luminosities of emission lines in the nebular spectra with the light curve peak magnitudes to reveal their power sources. In this work, we analyze 7 SNe IIb, as well as 2 SNe Ib (SN 2007Y and iPTF 13bvn) that show the H$\alpha$-like emission in their nebular spectra. The luminosity of the H$\alpha$-like emission shows a tight correlation with the light curve peak magnitude, sharing the same behavior with other nebular lines. This result indicates that the H$\alpha$-like emission is powered by the radiative decay of $^{56}$Co. The line flux is then expected to roughly follow the mass of the emitting layer. The variation expected from the diversity of the H-rich envelope mass among SNe IIb (reaching nearly to an order of magnitude) is however not seen, suggesting that it is most likely contributed predominantly by [N II], not by H$\alpha$. While further analysis is limited by the available sample size, we find a hint that SNe IIb with a double-peak light curve, which is interpreted as an outcome of the more extended and massive hydrogen envelope, tend to show excess in the luminosity of the H$\alpha$-like feature than those with a single-peak light curve. This behavior indicates possible additional contribution from H$\alpha$. Additionally, we also find a correlation between the [Ca II]/[O I] ratio and the post-maximum decline rate, indicating that the [Ca II]/[O I] can be used as a diagnostics for the progenitor mass. 
\end{abstract}
 
\keywords{
  line: formation --- supernova: core-collpase: nebular --- supernova: general}

%
%


\section{INTRODUCTION}
A star with zero-age main-sequence mass larger than 8 M$_{\odot}$ ends its life with a supernova (SN) explosion triggered by a collapse of its iron or oxygen-neon-magnesium core. Core collapse supernovae (CCSNe) are classified into type II SNe (SNe II, with a hydrogen envelope) and type I (SNe I, without a hydrogen envelope). Type I CCSNe are further divided into SNe Ib and Ic according to whether its helium envelope is retained. Observationally, Balmer lines shape optical spectra of SNe II, while they are (generally) not detected for SNe Ib and Ic. 

SNe IIb show spectroscopic properties intermediate between SNe II and Ib. Optical spectra of SNe IIb show strong hydrogen lines around the maximum light. As an SN IIb evolves, the hydrogen lines gradually fade away, eventually resembling to an SN Ib. The small amount of hydrogen retained at the time of the explosion is believed to be responsible for such a transition (\citealt{filippenko93}; \citealt{nomoto93}). The first identification of SN IIb was suggested for SN 1987K \citep{filippenko88}. The number of SNe IIb so far discovered is increasing, among which SN 1993J is the best-observed one. However, SN IIb is a relatively rare event, with the volumetric rate of about 12\% among SNe II (\citealt{li11}; \citealt{smith11}). 

In early phase, the emission from SNe IIb, including the hydrogen lines in their spectra, is powered by the radioactive decay chain $^{56}$Ni $\rightarrow$ $^{56}$Co $\rightarrow$ $^{56}$Fe. In nebular phase, they generally show an emission feature centered at $\sim 6500$ \AA. This may be contributed by H$\alpha$, but the line identification and the power source have not been robustly clarified. Indeed, for a fraction of SNe IIb, this feature further develops in later phases (e.g., after $\sim$300 days) and it is considered as a signature of strong interaction with circumstellar material (CSM)  (\citealt{patat95}, \citealt{matheson00a}, \citealt{maeda15}). The behavior is not always seen, and thus questions remain as for what are the line identification and power source for the H$\alpha$-like feature in nebular phase but at $< 300$ days (\citealt{maurer10}, \citealt{jerk15}, \citealt{maeda15}). This is related to the still-unresolved question of the mechanism of the envelope stripping toward SNe IIb, and toward SNe Ib and Ic in general (\citealt{grafener16}; \citealt{stancliffe09}; \citealt{ouchi17}).

Nebular line identification provides us with an opportunity to explore the properties of the entire ejecta structure from the core through the envelope. Although metal lines are unambiguously a result of the radioactive decay chain (\citealt{kozma98}; \citealt{houck96}), the origin of the H$\alpha$-like feature detected generally for SNe IIb already before $\sim 300$ days has not been clarified. For some SNe IIb, the H$\alpha$-like emission is relatively narrow at $< 300$ days, which is different than a flat and wide line profile predicted by the CSM interaction scenario (\citealt{cheva10}) as examplified by SN 2008ax (\citealt{tauben11}). \citet{cheva10} also argue that H$\alpha$ powered by shock-CSM interaction should be undetectable in nebular phase for relative compact objects (e.g., SN 2007Y), but this H$\alpha$-like structure still presents in their spectra. Note that while the luminosities of H$\alpha$-like structure in relatively early nebular phase are similar for SNe 1993J and 2008ax (\S 4 in this work, and \citealt{tauben11}), the mass-loss history affecting the interaction power is derived to be very different (\citealt{maund04}; \citealt{folatelli15}), questioning the shock-CSM interaction mechanism as a dominant power source at $< 300$ days.
 
Another possibility is that this emission (before $\sim 300$ days) is powered by the radioactive decay chain. \citet{patat95} argue that the mass of hydrogen envelop of SN IIb is not massive enough to produce such luminous emission through radioactivity, however, \citet{maurer10} suggest that if some amount of hydrogen is mixed into the helium layer, the radioactive decay chain could power the H$\alpha$ emission with a broad and boxy profile, although some assumptions in their scenario remain to be discussed. \citet{jerk15} include [N II] $\lambda \lambda$ 6548, 6583 in their synthetic spectral calculations, and find that the cooling within the He/N zone by [N II] can produce the H$\alpha$-like structure seen in SNe IIb in their sample. 

Although various scenarios have been proposed, a model-independent (observational and phenomenological) approach, especially based on the statistic behavior of this emission, is missing. As noted before, the early light curve of SNe IIb is powered by the radioactive decay chain, and the peak magnitude is correlated with the mass of $^{56}$Ni produced at the explosion \citep{lyman16}. However, the power provided by shock-CSM interaction has no direct link to the amount of radioactive decay isotopes, as it is mainly affected by the mass-loss history. Therefore, a combined analysis of early and late phase observations will provide clear diagnostic on the power source leading to this feature. In this work, we compare early and nebular observables to distinguish shock-CSM interaction and radioactive power scenarios. We also analyze the luminosity scatter level of this feature, in order to further constrain the identity of the feature.

Based on the results of this paper through the model-independent approach, we suggest this emission feature should be powered by the radioactive decay chain. It is more likely [N II] rather than H$\alpha$.

This paper is organized as follows. In \S 2, we first introduce the sample of SN IIb used in this paper, together with our methods of light curve and spectrum analyses. Relations between the observables and physical parameters are also briefly summarized in \S 2, which guides the interpretations in the following sections. In \S 3, we compare the observables in early phase with those in nebular phase. The discussion part is given in \S 4, and the paper closes in \S 5.

\section{Sample description}
In this work, we collect SNe IIb for which extensive photometric and spectroscopic data are available from early to nebular phases (up to $\sim$ 300 days after light curve peak in the $V$-band). However, the SN IIb is a rare event, and the faintness in the nebular phase is an impediment toward observation. These two factors make the sample size of SNe IIb in this study relatively small. In addition to SNe IIb, two SNe Ib, SN 2007Y and iPTF 13bvn, which show the H$\alpha$-like feature in their nebular spectra, are also included in our sample for comparison. In summary, our sample includes 7 SNe IIb and 2 SNe Ib, and 58 nebular spectra.

The analysis of this work is based on the spectra and photometric data compiled from the literature. Most of the spectral data are downloaded from Weizmann Interactive Supernova data REPository\footnote{https://wiserep.weizmann.ac.il/} (WISeREP, \citealt{yaron12}). The sources of the photometric data, distances and reddening adopted in this work are listed in Table~\ref{tab:basic_data}. The spectral sequence is listed in Table~\ref{tab:spec_log}. The date when the $V$-band maximum is reached is derived from low-degree polynomial fitting, and it is employed as the baseline of the photometric or spectral phases throughout this work.

\begin{table}[!b]
\begin{center}
\caption{Data and references for SNe in this work.}
\label{tab:basic_data}
\begin{tabular}{lcccc}
\hline
Object&$E(B - V)_{\rm total}$&Distance module& LC source\\
&(mag)&(mag)&\\
\hline
SN 1993J &0.19 (1) &27.81 $\pm$ 0.12 (1)& 2\\ 
SN 2003bg&0.02 (3) &31.68 $\pm$ 0.14 (3)& 3\\ 
SN 2007Y &0.11 (4) &31.36 $\pm$ 0.14 (4)& 4\\ 
SN 2008ax&0.40 (5) &29.92 $\pm$ 0.29 (5)& 5, 6\\ 
SN 2011dh&0.07 (7) &29.46 $\pm$ 0.10 (7)& 7\\ 
SN 2011hs&0.17 (8) &31.85 $\pm$ 0.15 (8)& 8\\ 
SN 2011fu&0.10 (9) &34.36 $\pm$ 0.15 (9)& 9\\ 
SN 2013df&0.10 (10)&31.65 $\pm$ 0.30 (10)& 10\\ 
iPTF13bvn&0.12 (11)&32.14 $\pm$ 0.20 (12)& 11\\
\hline%
\end{tabular}
\begin{tablenotes}[!b]
      \small
      \item \textbf{\textit{References.}} (1) \citet{matheson00}; (2) \citet{richmond96}; (3) \citet{hamuy09}; (4) \citet{strit09}; (5) \citet{tauben11}; (6) \citet{tsve09}; (7) \citet{ergon14}; (8) \citet{bufano14}; (9) \citet{morales15}; (10) \citet{morales14}; (11) \citet{fremling16}; (12) \citet{tully13}.
    \end{tablenotes}
\end{center}
\end{table}

\begin{table}[!b]
  \begin{threeparttable}
\caption{Spectrum sequence in this work}
\label{tab:spec_log}
\begin{tabular}{lccc}
\hline
Object&Date&Phase \tnote{a}& Ref.\\
 \hline
1993J &1993/09/14&149 &\citet{barbon95}\\ 
      &1993/09/15&150 &\citet{barbon95}\\
      &1993/09/20&155 &\citet{jerk15}\\
      &1993/10/19&184 &\citet{barbon95}\\
      &1993/11/07&203 &\citet{jerk15}\\
      &1993/11/15&211 &\citet{barbon95}\\
      &1993/11/19&215 &\citet{barbon95}\\
      &1993/12/08&234 &\citet{barbon95}\\
      &1993/12/17&243 &\citet{jerk15}\\
      &1994/01/05&262 &\citet{jerk15}\\
      &1994/01/21&278 &\citet{barbon95}\\
      &1994/02/17&305 &\citet{jerk15}\\
      &1994/03/09&325 &\citet{barbon95}\\
      &1994/03/10&326 &\citet{barbon95}\\
      &1994/03/25&341 &\citet{barbon95}\\
      &1994/03/30&346 &\citet{barbon95}\\
      &1994/05/14&391 &\citet{modjaz14}\\
      &1994/05/17&394 &\citet{jerk15}\\
      &1994/06/12&420 &\citet{modjaz14}\\
2003bg&2003/08/20&153 &\citet{hamuy09}\\ 
      &2003/09/18&182 &\citet{hamuy09}\\
      &2003/11/16&241 &\citet{hamuy09}\\
      &2003/11/29&254 &\citet{hamuy09}\\
      &2003/12/16&271 &\citet{hamuy09}\\
      &2003/12/23&278 &\citet{hamuy09}\\     
2007Y &2007/09/22&200 &\citet{strit09}\\ 
      &2007/10/21&229 &\citet{strit09}\\ 
      &2007/11/30&269 &\citet{strit09}\\ 
2008ax&2008/07/24&122&\citet{tauben11}\\ 
      &2008/07/30&128&\citet{tauben11}\\
      &2008/08/01&130&\citet{mili10}\\
      &2008/11/24&245&\citet{tauben11}\\
      &2008/12/08&259&\citet{tauben11}\\
      &2009/01/25&308&\citet{modjaz14}\\
      &2009/02/25&338&\citet{tauben11}\\
      &2009/04/22&394&\citet{mili10}\\      
2011dh&2011/12/18&182&\citet{shivvers13}\\
      &2011/12/19&183&\citet{ergon15}\\
      &2011/12/24&188&\citet{shivvers13}\\
      &2012/01/25&220&\citet{ergon15}\\
      &2012/02/23&249&\citet{shivvers13}\\ 
      &2012/03/18&273&\citet{ergon15}\\
      &2012/05/24&340&\citet{ergon15}\\
      &2012/07/19&396&\citet{ergon15}\\ 
2011hs&2012/05/01&160 &\citet{bufano14}\\ 
      &2012/06/21&211 &\citet{bufano14}\\ 
      &2012/06/03&213 &\citet{bufano14}\\ 
2011fu&2012/02/22&133&\citet{morales15}\\ 
      &2012/07/20&282&\citet{morales15}\\ 
2013df&2013/11/08&135&\citet{maeda15}\\ 
      &2013/12/05&162&\citet{morales14}\\ 
      &2013/12/21&178&\citet{maeda15}\\ 
      &2014/02/04&223&\citet{morales14}\\ 
      &2014/05/06&314&\citet{maeda15}\\ 
      &2015/02/21&602&\citet{maeda15}\\ 
13bvn &2014/02/21&234&\citet{fremling16}\\
      &2014/05/28&329&\citet{fremling16}\\
      &2014/06/26&359&\citet{fremling16}\\
 \hline%
 \end{tabular}
 \begin{tablenotes}
            \item[a] Phase relative to $V$-band peak listed in Table~\ref{tab:lc_para}.
         
        \end{tablenotes}
     \end{threeparttable}
\end{table}

\subsection{Photometric data}

In this section, we describe our light curve analysis method. We also briefly summarize the relation between the observables and physical properties of CCSNe.

\subsubsection{Light curve in early phase}
\begin{figure}[!t]
\includegraphics[width = 8cm]{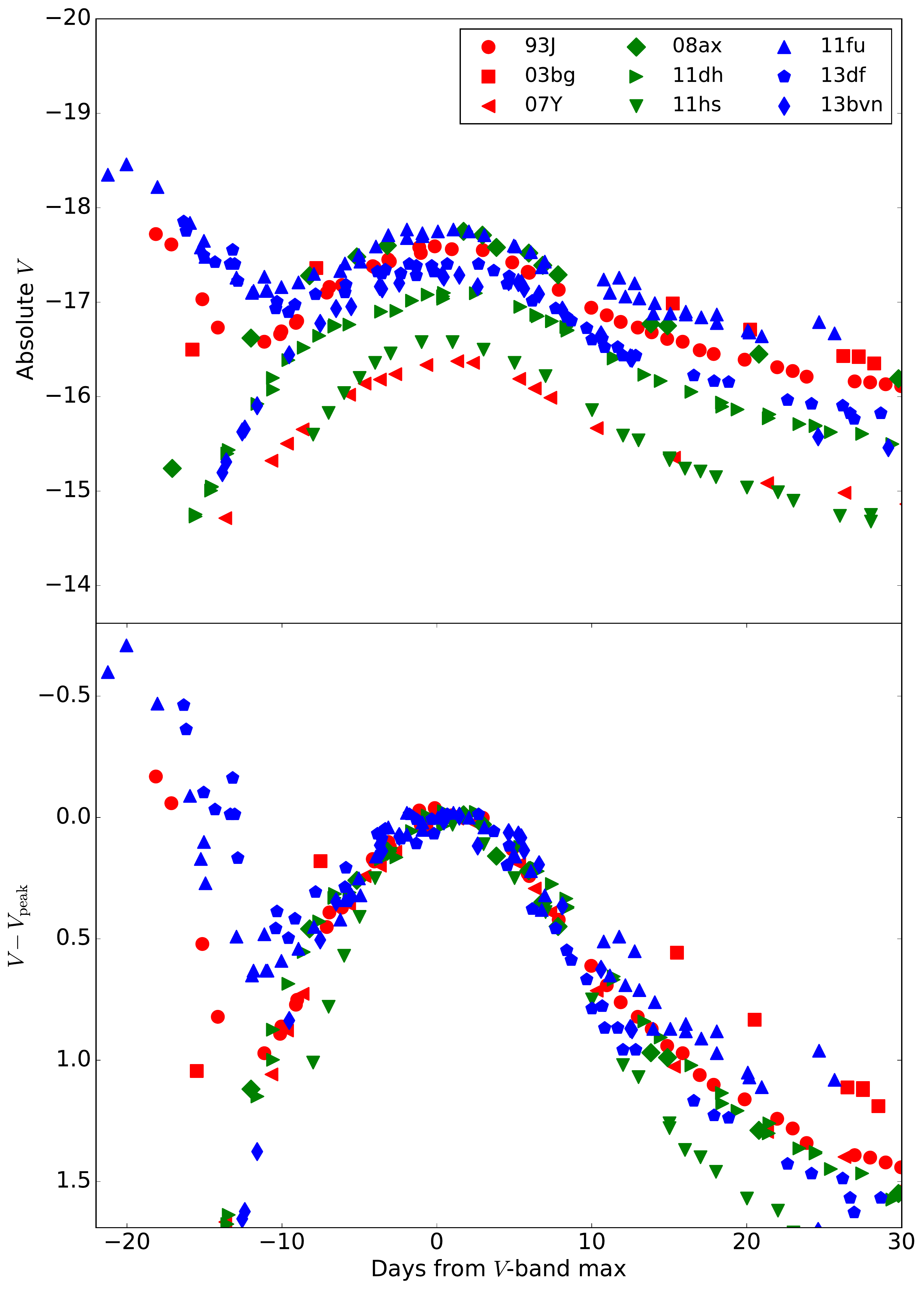}
\centering
\caption{The $V$-band light curves of 7 SNe IIb and 2 SN Ib studied in this work. In the top panel, the absolute magnitude scale is adopted. In the bottom panel, the magnitudes are normalized by the peak magnitudes.}
\label{fig:lc_div}
\vspace{4mm}
\end{figure}

The photometric data of SNe used in this work are compiled from the literature. After correcting for extinction and distance moduli listed in Table~\ref{tab:basic_data}, Figure~\ref{fig:lc_div} shows the $V$-band light curves, which highlight the similarity and diversity among the sample. The scatter in the peak magnitude can reach to $\sim$ 1.5 mag, and the decline rates after the maximum brightness also vary. Differences in the peak magnitude and the shape of the light curves imply some diversities in the explosion parameters.

The interpretation between the observables to the explosion parameters can be constructed in a simple manner (\citealt{valenti08}; \citealt{lyman16}). As shown in Figure 5 of \citet{lyman16}, the peak bolometric magnitude is tightly correlated with the mass of $^{56}$Ni ($M_{\rm Ni}$) ejected by the explosion. \citet{morales16} also find a correlation between the $R$-band peak magnitude and $M_{\rm Ni}$ for SNe IIb. In principle, we could (roughly) convert the bolometric peak magnitude to the $^{56}$Ni mass using an analytical model \citep{arnett82}. However, the absolute value of the $^{56}$Ni mass is not important in this work. Therefore we use the relative values of the $V$-band peak magnitudes as a representative observable to be connected to the variation of the mass of $^{56}$Ni among the sample of SNe IIb (and SN Ib) to reduce uncertainty from analytical models. The mass of $^{56}$Ni is then characterized by the $V$-band peak magnitudes as
\begin{equation}
  {\rm log}_{10}M_{\rm Ni} \sim -0.4 \times V_{\rm peak} + {\rm constant} 、 ,
 \label{eq:Vpeak_Ni}
\end{equation}   
where $M_{\rm Ni}$ is the mass of $^{56}$Ni and $V_{\rm peak}$ is the $V$-band peak magnitude. Since we are interested only in the relative values of $M_{\rm Ni}$, constant in Equation~\ref{eq:Vpeak_Ni} can be taken as a given arbitrary value. 

The light curve parameters also contain information of the ejecta mass. For simplicity, we assume luminosity of an SN decreases as
\begin{equation}
  L(t) = L_{\rm peak}\times {\rm exp}\left(-\frac{t}{t_0}\right) \ , 
 \label{eq:Lum_t}
\end{equation}  
$t_0$ is the characteristic time scale for the luminosity decrease and $L(t)$ is the luminosity at time $t$ , where $t = 0$ represents $V-$ band peak. In magnitude scale, 
\begin{equation}
  M(t) = M_{peak} + 1.08 \times \frac{t}{t_0} \ ,
 \label{eq:Mag_t}
\end{equation}
where $M(t)$ is the magnitude at time $t$ after peak is reached.

Analogous to SNe Ia, we characterize the light curve width by $\Delta m_{15}$, i.e. a change in the magnitude for $15$ days after the light curve peak. Again, we choose $V$-band in our analysis. Therefore, we have
\begin{equation}
  \Delta m_{15} = \frac{18.2}{t_0} \propto t^{-1}_{0} \ .
 \label{eq:dm15_tau}
\end{equation}

The width of the light curve carries information on the ejecta mass and the kinematic energy. Equation (1) of \citet{valenti08} gives
\begin{equation}
  w_{lc} \propto \frac{M^{3/4}_{\rm ejecta}}{E^{1/4}_{\rm K}} \ ,
 \label{eq:tau_Mej}
\end{equation}
where $w_{lc}$ is the width of light curve, $M_{\rm ejecta}$ is the ejecta mass and $E_{\rm K}$ is the kinematic energy (see also \citealt{arnett82}). Since the scatter in the expansion velocities of SN IIb is relatively small (8300 $\pm$ 750 km/s, see \citealt{lyman16}). we have

\begin{equation}
    {\rm log}M_{\rm ejecta} \sim -2\times {\rm log}\Delta m_{15} \ .
 \label{eq:dm15_Mej}
\end{equation}

\begin{figure*}[!t]
\includegraphics[width = 18cm]{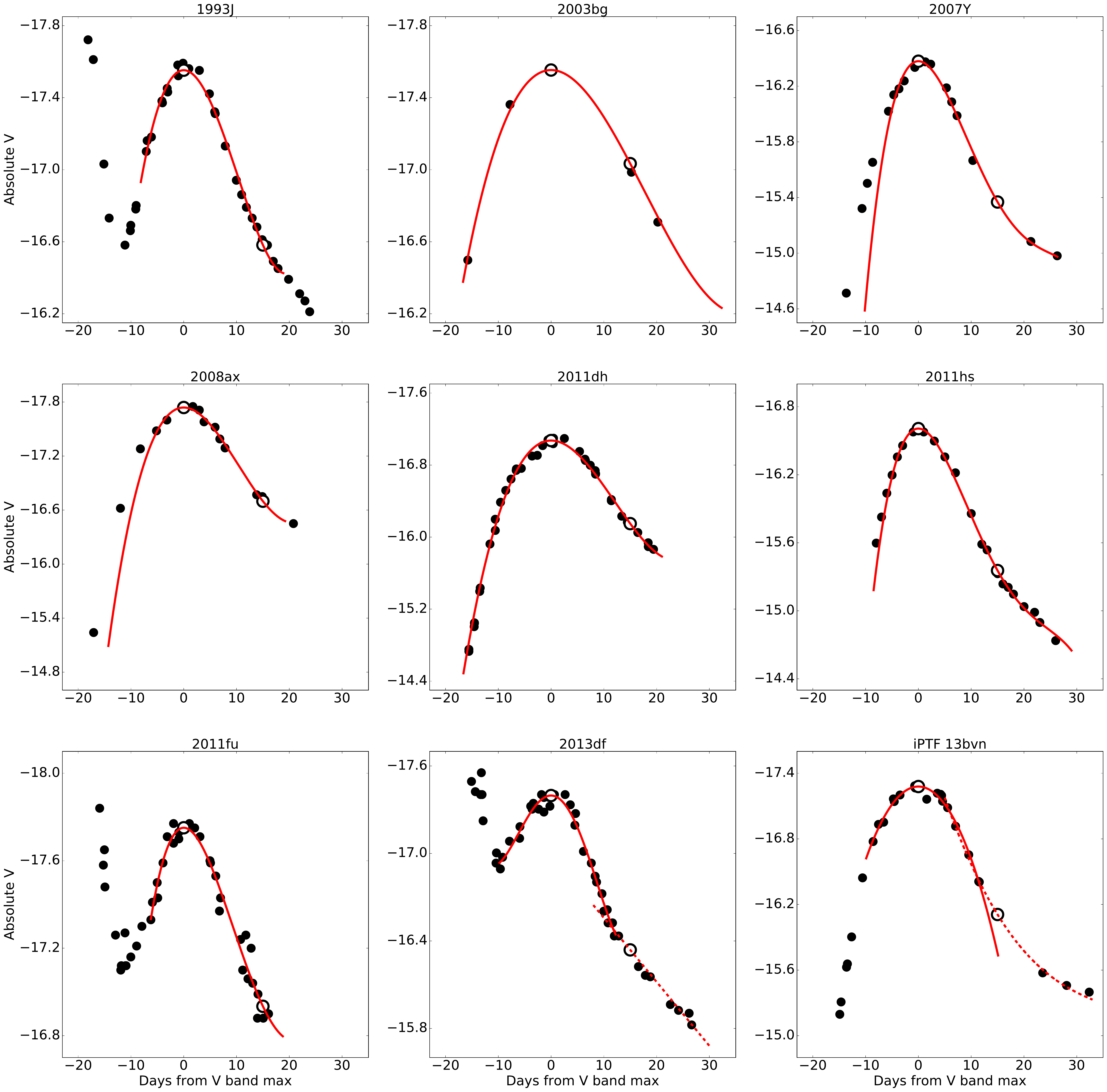}
\centering
\caption{The $V$-band light curves corrected for extinction and distance moduli (filled points). The red solid line in each panel is the fitting result by a low-degree polynomial function. The derived peak magnitude and the magnitude at $15$ days after the maximum are denoted by the open circles.}
\label{fig:lc_fit}
\end{figure*}

To derive the characteristic light curve parameters, we fit the light curve data by a low-degree polynomial function. For SNe IIb, the typical maximum date in the $V$ band is $\sim 20$ days after the explosion. As show in Figure~\ref{fig:lc_div}, some SNe IIb show a light curve with two peaks. The power source of the first peak is the thermal energy deposited at the explosion (so called `cooling emission', see \citealt{arnett80}), and that of the second peak is the radioactive decay chain of $^{56}$Ni and $^{56}$Co. To restrict our analysis to the radioactive-powered peak, only data points at $5 \sim 30$ days after the explosion are fitted. An additional polynomial fit after the (second) peak is also applied to estimate $\Delta m_{15}$ whenever necessary. The results are shown in Figure~\ref{fig:lc_fit}. A low-degree polynomial fit provides reasonable fitting for most objects, although for some objects (SNe 2007Y and 2008ax), the fitting results slightly deviate from the observations at very early phase. This inconsistency can be remedied by raising the degree of the polynomial function. However, our analysis on the early light curve is only restricted to the phase of the maximum and $15$ days after. To avoid introducing additional uncertainty by enlarging the parameter space, we restrict our fitting by using a polynomial function with $3 \sim 5$ degrees and the inconsistency at very early phase is neglected. The peak magnitude and that at $15$ days after the maximum are labeled by the open circles in Figure~\ref{fig:lc_fit}. 

For simplicity, we neglect uncertainty in the peak magnitude, since the photometric error around peak is small. The errors in distance and extinction are not included in Figures 1 \& 2, while they are added in the subsequence analyses (note however that it would not introduce a random scatter in most of the analyses in this paper, as these errors enter into the early-phase and late-phase fluxes in the same manner). Uncertainty in $\Delta m_{15}$ is negligible in most of the sample as it does not depend on the estimate on the distance and extinction. The main error budget in $\Delta m_{15}$ comes from uncertainty in the peak date, but this is constrained as $\pm$ 1 days. The largest uncertainty in $\Delta m_{15}$ is found for SN 2003bg, which lacks well-sampled photometric data around the maximum, and the error in the peak date is as large as $\pm $ 2 days \citep{hamuy09}. We calculate the magnitudes at 15 $\pm$ 1 (or $\pm$ 2 for SN 2003bg) days after the maximum brightness, and the deviation is defined to be the uncertainty of $\Delta m_{15}$. The light curve parameters and uncertainties adopted in this work are listed in Table~\ref{tab:lc_para}.

\begin{table}[!b]
\begin{threeparttable}
\caption{Light curve parameters}
\label{tab:lc_para}
\begin{tabular}{lccc}
\hline
Object&$t_{\rm peak}$ (JD - 2400000)\tnote{a}&$V_{\rm peak}$&$\Delta m_{15}$\\
 \hline
1993J &49094.8&-17.55&0.97 $\pm$ 0.06\\ 
2003bg&52718.3&-17.56&0.52 $\pm$ 0.11\\ 
2007Y &54164.2&-16.38&1.01 $\pm$ 0.06\\
2008ax&54549.6&-17.74&1.04 $\pm$ 0.07\\
2011dh&55732.0&-17.07&0.92 $\pm$ 0.08\\
2011hs&55888.5&-16.61&1.25 $\pm$ 0.08\\
2011fu&55846.4&-17.75&0.82 $\pm$ 0.05\\
2013df&56469.8&-17.40&1.06 $\pm$ 0.04\\
13bvn &56475.1&-17.28&1.17 $\pm$ 0.08\\
\hline%
 \end{tabular}
 \begin{tablenotes}
            \item[a] The peak of light curve is relatively well-constrained for most objects, and we assume that the typical uncertainty is 1 day. For SN 2003bg, the light curve sampling is coarse and the uncertainty of peak date is assumed to be 2 days.
         
        \end{tablenotes}
\end{threeparttable}
\end{table}

\subsubsection{Light curve in nebular phase}
For long-slit spectroscopy, absolute flux calibration can be difficult, and spectra downloaded from WISeREP may previously be normalized only with spectroscopic standard stars. Therefore, we use photometric data to further calibrate the absolute flux of the nebular spectra in this study. We choose $R-$band magnitude for the calibration, because it is generally better sampled than $V$-band magnitudes in nebular phase, and the strong nebular lines of interest in this work fall within the wavelength range of this bandpass. However, the $R$-band light curves of SN 2007Y and iPTF 13bvn are not available. We thus use $r$-band photometric data to calibrate the nebular spectra of these SNe. For spectra which do not cover the wavelength range of $R$-band, we alternately use $V$-band photometric data to anchor their fluxes.

The emission in nebular phase is mainly powered by the radioactive decay chain. The light curve therefore behaves quasi-linearly in the magnitude scale as a function of time \citep{maeda03}. Later on, the behavior may be complicated by increasing contribution from the positron deposition \citep{cappe97} or shock-CSM interaction. Therefore a linear or quadratic fit to the light curve is applied at $> 60$ days, in order to estimate $R$-band magnitude for each spectrum. Following \citet{bessell12}, we calculate the spectroscopic magnitude for a given spectrum by convolving it with the filter function. The deviation of the the spectroscopic magnitude from observed photometry is then used to anchor the flux scale.

\subsection{Nebular spectrum}
In early phase, only the outermost layer of SN ejecta is observed. As it evolves, the ejecta becomes more transparent, and a deeper region is exposed. Several months after explosion, an SN enters into nebular phase, where its spectrum is dominated by forbidden lines superposed on a faint continuum. The luminosities and structures of the nebular emission lines are useful tracers of the properties of the ejecta (\citealt{maeda08}; \citealt{modjaz08}; \citealt{tauben09}).

In this work, we compare early and nebular phase observables to reveal the power source(s) of the H$\alpha$-like structure which appears in nebular spectra of SN IIb and some SNe Ib. The well-studied nebular lines, including [O I] doublet, [Ca II] doublet and Na I D are also studied for comparison. To this end, nebular spectra are pre-processed as follows. First, absolute fluxes are calibrated to match to the photometric data. After that, the spectra are de-reddened with the Cardelli extinction law (\citealt{cardelli89}), assuming $R_V$ = 3.1. For simplicity, the total extinction is applied to the observed wavelength. This is not exact for the host extinction, but the effect is negligible for our low-redshift samples. Finally, redshifts are corrected.

\begin{figure*}[!t]
\includegraphics[width = 18cm]{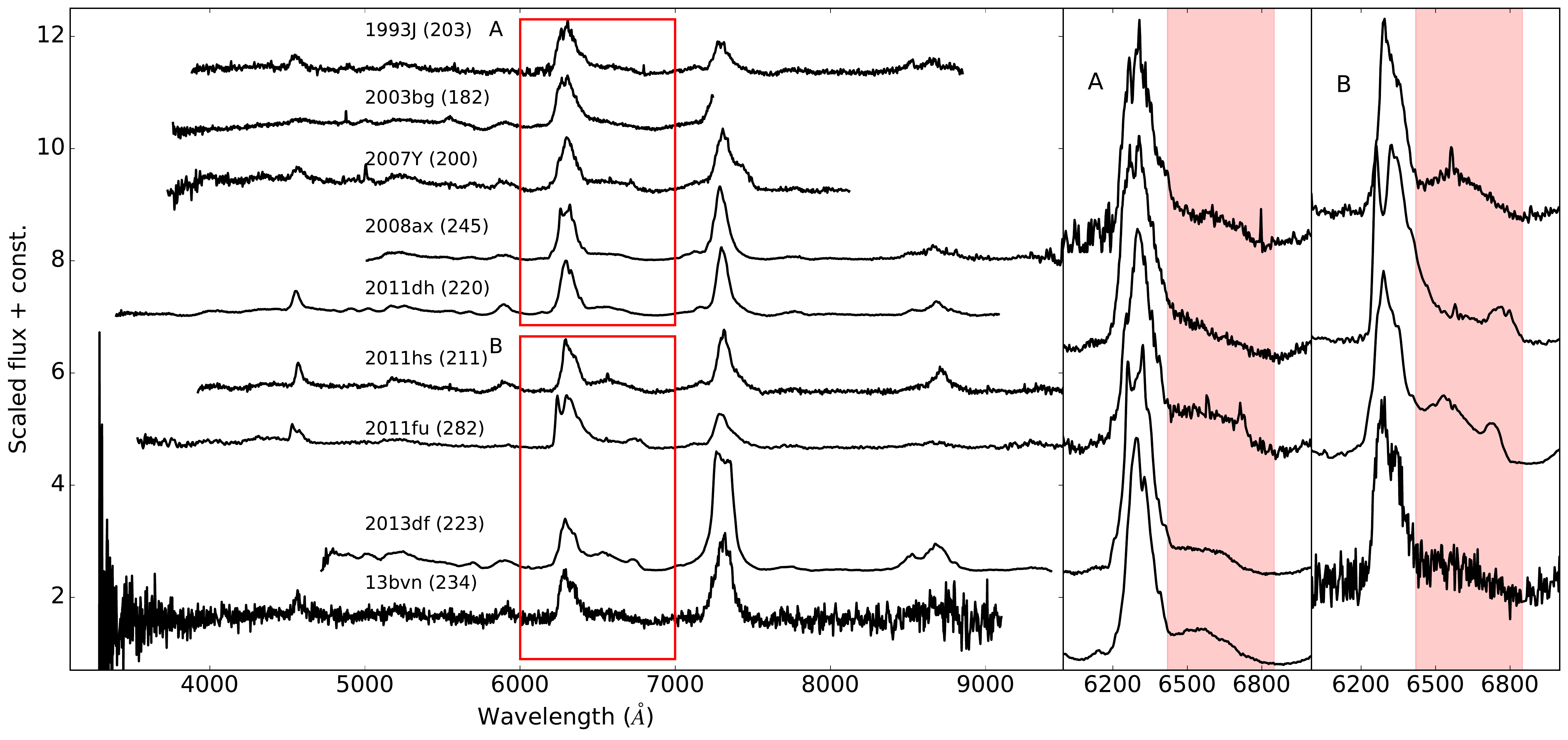}
\centering
\caption{The spectra of SNe IIb/Ib at $\sim$ 200 days after the peak magnitude is reached. All spectra are shifted to the rest wavelength and phases relative to $V$-band maximum are listed in parenthesis. In the right panels, the wavelength region for the emission complex including the [O I] and H$\alpha$-like feature is shown.}
\label{fig:spec_ill}
\end{figure*}

Figure~\ref{fig:spec_ill} shows the overall spectra of our sample at $\sim$ 200 days after the light curve peak is reached. The color region illustrates the H$\alpha$-like feature.

\subsubsection{[O I] $\lambda \lambda$ 6300, 6363 and H$\alpha$ like structure}
The [O I] doublet and the H$\alpha$-like structure together form an emission complex. To separate these two components, the first step is to remove local continuum from the spectrum. For an emission-free wavelength regime, determining the local continuum is straightforward. We first (slightly) smooth a spectrum by convolving it with a Gaussian kernel, then we find the local minima within a wavelength range of a few $100$ \AA~on both sides of this emission complex. A strait line connecting these two minima is defined to be the local continuum.

\begin{figure*}[!t]
\includegraphics[width = 15cm]{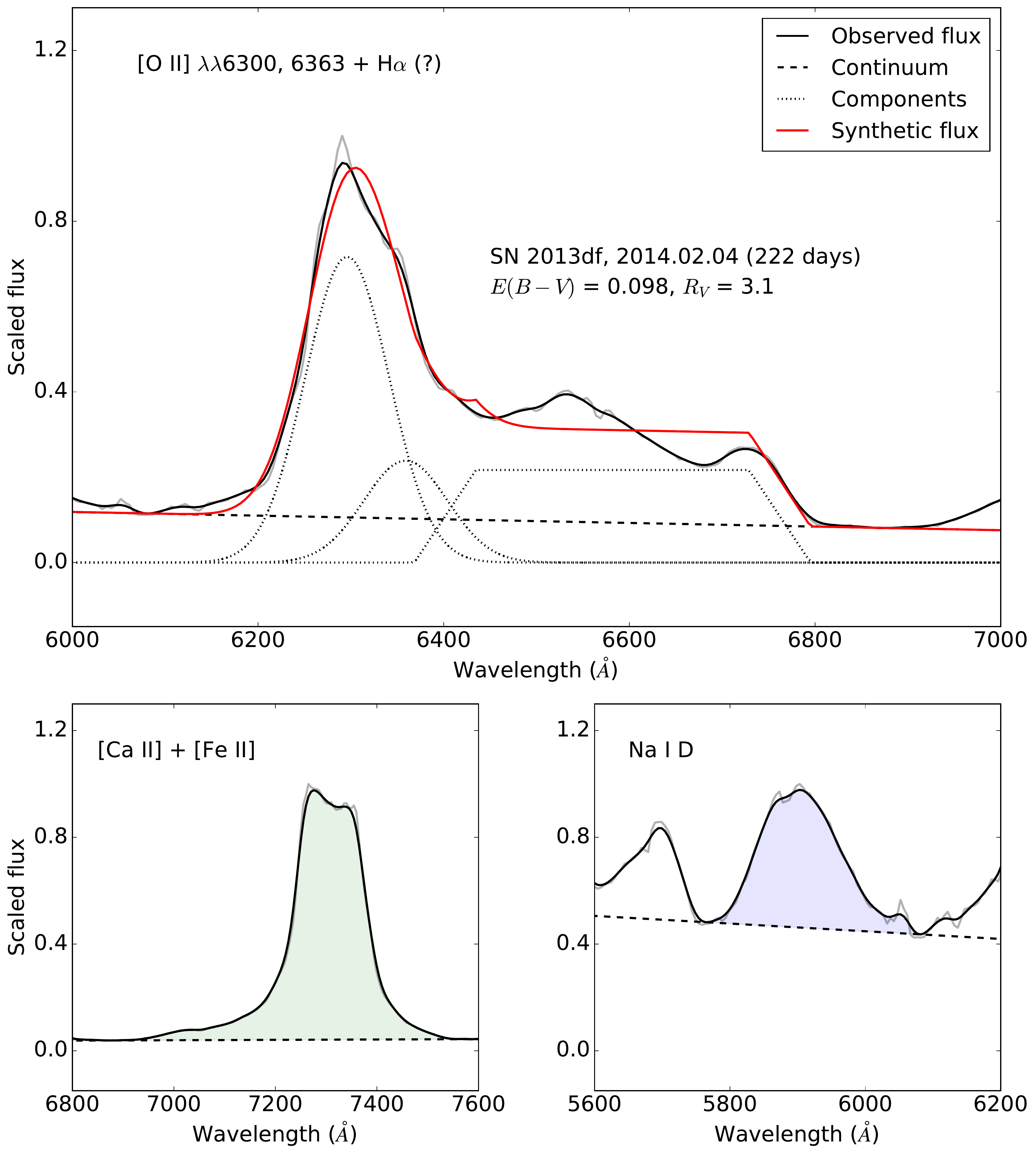}
\centering
\caption{An example of the line decomposition. Top panel: the black solid line is the observed spectrum, and the red solid line is the composed spectrum from the fits to the different components. The local continuum and different components are presented by black dotted and dashed lines, respectively. Bottom panels: illustrations for the continuum determinations and integral fluxes of Na I and [Ca II].}
\label{fig:example_decomp}
\vspace{4mm}
\end{figure*}

The H$\alpha$-like structure is de-blended from [O I] $\lambda \lambda$ 6300, 6363 after local continuum is removed. The line profile of the [O I] is assumed to be double-Gaussian. The centers of these two Gaussian functions are fixed at 6300, 6363 \AA, and they are assumed to have the same velocity structure (therefore, same $\sigma$ in Gaussian profile). To avoid contamination from the H$\alpha$-like structure, only the blue part of the doublet is fitted. We tested two options for the ratio of the doublet peaks, either leaving it as a free parameter or fixing it to be 3:1 expected in the optically thin limit. We find virtually no difference in the fitted flux of the [O I], and further in the subtraction of this feature from the spectra. Therefore, in the subsequent analysis, the line ratio is fixed to be 3:1, which is generally applicable to SNe IIb/Ib/Ic with a few exceptions (\citealt{elmham11}, \citealt{jerk15}). After removing the local continuum and the fitted [O I] profile, the residual flux represents the H$\alpha$-like emission. It is then fitted with a flat-topped profile. An example of this fitting procedure is presented in the top panel of Figure~\ref{fig:example_decomp}. 

\begin{figure}[!t]
\includegraphics[width = 8cm]{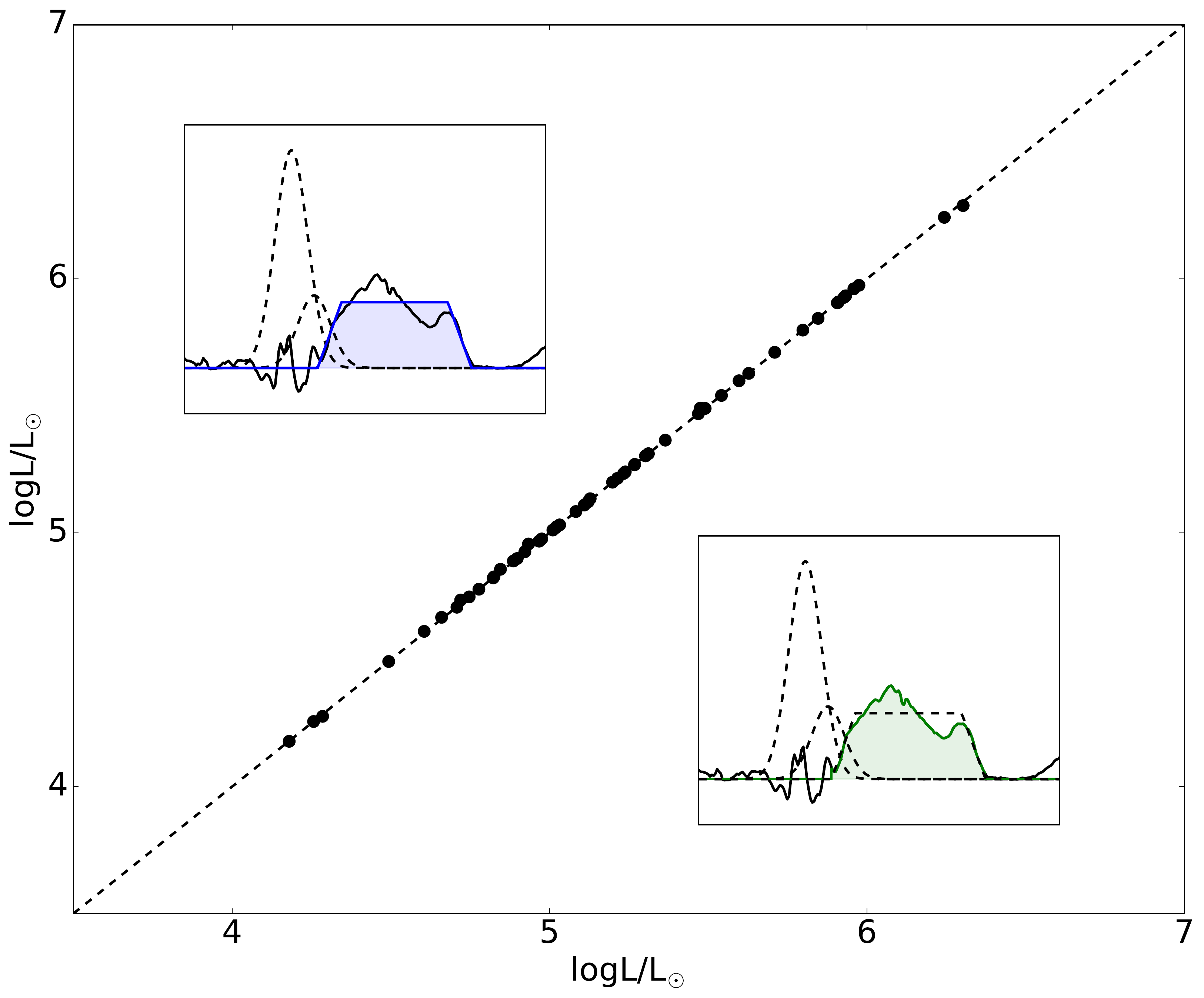}
\centering
\caption{A comparison between the results of the two-fitting strategies to the H$\alpha$-like fluxes in the nebular spectra. Illustrations of these two fitting strategies are presented, where the integrations of the colored areas are adopted as the flux of the H$\alpha$-like feature.}
\label{fig:example_fit}
\vspace{4mm}
\end{figure}

However, the power source of this emission line is a topic to investigate in this paper. Assuming it to be flat-topped, which is the character expected from the shock-CSM interaction scenario, indeed conflicts with the purpose of this work. The flux of this emission line can also be calculated by integrating over the spectrum after the local continuum and the fitted oxygen doublets are subtracted, where upper and lower limits are the zero-flux points in the fitted flat-topped profile. Figure~\ref{fig:example_fit} illustrates and compares these two strategies for determining the flux of the H$\alpha$-like feature, showing that they give mutually consistent results. In what follows, we will thus adopt the integral flux. This method does not assume the center wavelength and the line profile, so it is indeed a better and more model-free approach to tackle to the origin and power source of this particular emission feature than assuming a flat-topped profile.

\subsubsection{[Ca II] $\lambda \lambda$ 7291, 7324 and Na I D}
[Ca II] $\lambda \lambda$ 7291, 7324 and [Fe II] $\lambda$ 7155 are difficult to de-blend. This complex, together with Na I D, will be employed as tracers of the nebular lines powered by the radioactive decay.

The first step is to remove the local continuum. As is the same for [O I] and the H$\alpha$-like feature, the local continuum is assumed to be the line determined by the minima at both sides of the emission feature. After subtracting it, the flux is integrated, where the upper and lower limits are given by the two minima. 
Although a line decomposition method for the [Ca II]/[Fe II] complex has been discussed in \cite{terr16}, the detailed line profile of each emission is not a topic to be investigated in this paper. In any case, [Ca II] dominates this feature, and further subtraction of the minor contribution from [Fe II] to the flux of this complex does not affect our conclusions and arguments (see, e.g., Figure~\ref{fig:example_decomp} for the wavelength dependence). To avoid introducing uncertainty from line decomposition, we only discuss the sum of the [Ca II] and [Fe II] emissions. An illustration of the integrations of the fluxes of the [Ca II]/[Fe II] complex and Na I D is presented in the bottom panels of Figure~\ref{fig:example_decomp}. 

Uncertainties of the line luminosities mainly come from uncertainty in the photometric magnitude used for the flux calibration, except for the errors in the distance and extinction. In this work, a typical error of nebular photometric magnitude is estimated to be 0.1 mag, and it is quadratically added with uncertainty from the magnitude estimation (see \S2.2) which includes the uncertainty in the distance and the extinction. Uncertainty in the magnitude is then converted to the luminosity scale. We also include uncertainty from the continuum determination, which is conservatively estimated to be 10\% in this work. Note that the uncertainties in distance and extinction cancel out in comparing the flux scales in early and late phases. 

\subsection{Line luminosity evolution}
\begin{figure*}[!t]
\includegraphics[width = 18cm]{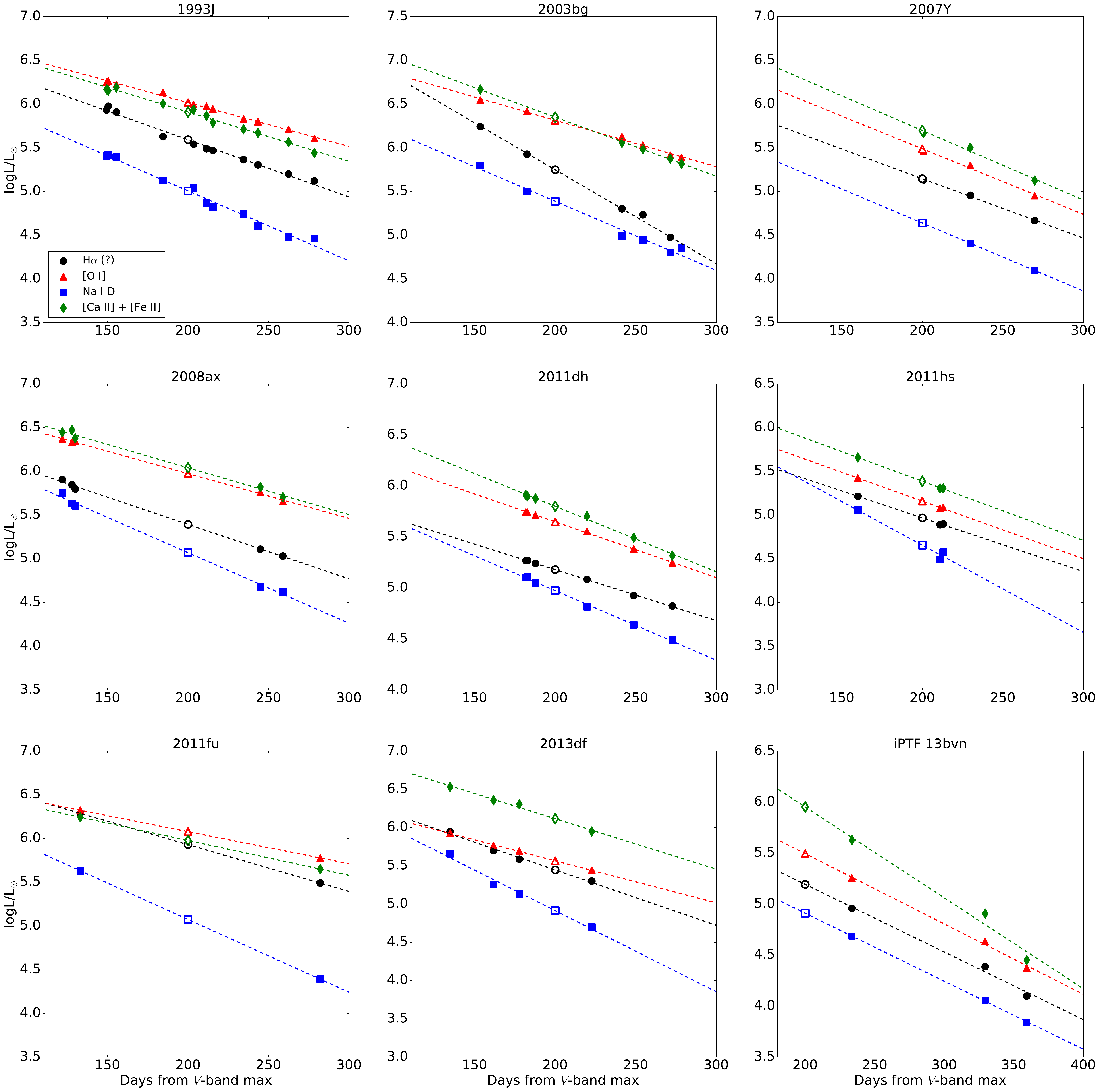}
\centering
\caption{Evolution of the luminosities of the different nebular lines, which are labeled by different colors and markers. Dashed lines are the best-fitted lines. The open markers are the estimated luminosities at 200 days after the $V$-band maximum.}
\label{fig:line_evo}
\vspace{4mm}
\end{figure*}

Figure~\ref{fig:line_evo} shows the luminosity evolution, where the H$\alpha$-like structure and other nebular lines decline linearly (in logarithmic scale) at $<$ 300 days. The dashed lines give the best linear fits, using luminosities only at $<$ 300 days for most objects (except for iPTF 13bvn, which is fitted using all available spectroscopic data). In comparing the early and late-time fluxes, line luminosities are all placed at the same phase ($200$ days after $V$-band maximum). The estimated luminosities are labeled by open markers.    

To calculate the uncertainty of the estimated line luminosities, we perform 10$^3$ Monte Carlo simulations. In each simulation, the line luminosities are perturbed by Gaussian error reported in \S 2.2. Then we perform a linear fit and estimate the luminosity at $t = 200$ days. The standard deviation from these simulations is taken as the uncertainty of the luminosities at $200$ days. The luminosities and their uncertainties thus obtained are listed in Table~\ref{tab:neb_lum}.
                                                                                                                                                                                                                                                                                           
\section{Results}
\begin{figure*}[!t]
\includegraphics[width = 15cm]{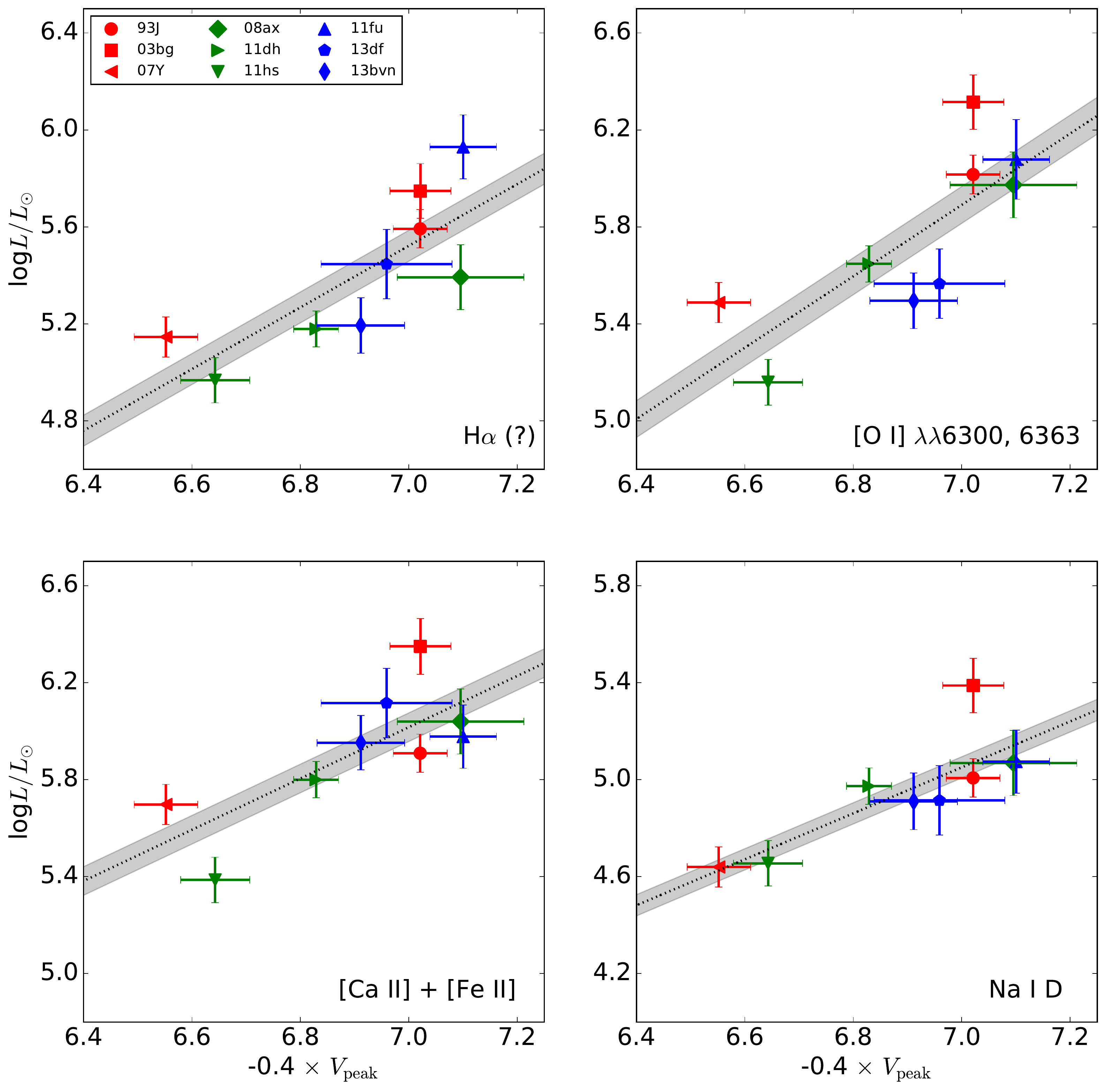}
\centering
\caption{ $V$-band peak magnitudes (representing relative log$M_{\rm Ni}$) versus the luminosities of the nebular lines in logarithmic scale, at $200$ days after the $V$-band maximum. Error in distance is included. In each plot, the dashed line is the best-fit result and shadow region is the standard deviation of the fitting. The slopes are listed in Table~\ref{tab:neb_lum}.}
\label{fig:Ni_luma}
\vspace{4mm}
\end{figure*}

In Figure~\ref{fig:Ni_luma}, the luminosities of the nebular lines (in logarithmic scale) are plotted against $-0.4~ \times V_{\rm peak}$, which represents a relative value of $M_{\rm Ni}$ produced in the explosion. A clear correlation is seen for all the emission lines. Pearson correlation coefficients are $r = 0.79$, $0.78$, $0.76$, $0.82$ for the H$\alpha$-like structure, [O I], [Ca II]/[Fe II] complex, and Na I D, respectively. For [Ca II]/[Fe II], Na ID, and [O I], this correlation is expected, given the robust identity of these lines as radioactive-decay powered metal lines (\citealt{houck96}; \citealt{kozma98}; \citealt{jerk15}). However, the correlation between the $^{56}$Ni mass and the luminosity of the H$\alpha$-like structure is not readily foreseen. Indeed, in the shock-CSM interaction scenario, the flux of this feature is expected to be correlated with the mass-loss history before the explosion, leaving no direct link to the $^{56}$Ni mass. The correlation between the nebular line luminosities and the peak magnitude thus implies that they share the same power source, i.e. the radioactive decay chain, with other metal lines. The slopes in the fits are listed in Table~\ref{tab:neb_lum}.

\begin{table*}[!t]
\begin{center}
  \begin{threeparttable}
\caption{Nebular luminosities in this work}
\label{tab:neb_lum}
\begin{tabular}{lcccccccccc}
\hline
Line\tnote{a}&1993J&2003bg&2007Y&2008ax&2011dh&2011hs&2011fu&2013df&13ibvn&slope\\
\hline
H$\alpha$ (?)                             &5.59&5.75&5.15&5.39&5.18&4.97&5.93&5.46&5.19&1.27 $\pm$ 0.37\\ 
$[$O I$]$ $\lambda \lambda$ 6300, 6363  &6.02&6.31&5.49&5.97&5.65&5.16&6.08&5.58&5.49&1.47 $\pm$ 0.44\\  
$[$Ca II$]$ + [Fe II]                       &5.91&6.35&5.70&6.04&5.80&5.39&5.98&6.13&5.95&1.06 $\pm$ 0.34\\
Na I D                                    &5.01&5.39&4.64&5.07&4.97&4.65&5.07&4.92&4.91&0.95 $\pm$ 0.25\\ 
Uncertainty\tnote{b}                      &0.06&0.10&0.07&0.07&0.06&0.07&0.12&0.08&0.09&-\\
\hline
\end{tabular}
\begin{tablenotes}
    \item \textbf{\textit{Notes.}} Line fluxes are calculated as Section 2.2 and distances are listed in Table~\ref{tab:basic_data}. The reported error includes error of photometric data, uncertainty from background determination (which is estimated to be 10$\%$, corresponds to $\sim$ 0.04 dex in logarithm scale), uncertainty from line fitting procedure and uncertainty from the estimation of luminosity described in Section 3. Error in distance listed in Table~\ref{tab:basic_data} is quadratically added whenever necessary.
     \item[a] Units: logL/L$_{\odot}$, where L$_{\odot}$ is fixed to 3.842 $\times$ 10$^{33}$ erg s$^{-1}$.
     \item[b] Uncertainty from photometric adopted for flux calibration is dominated, therefore different emission lines have almost same uncertainty.
\end{tablenotes}
   \end{threeparttable}
\end{center}
\vspace{4mm}
\end{table*}

\subsection{Effect of the gamma-ray energy deposition}
Although the correlations in Figure~\ref{fig:Ni_luma} imply that the nebular lines are powered by the radioactive decay chain, the luminosity scatter is relatively large. We now investigate whether such a scatter would be mainly originated in a different amount of the gamma ray energy deposition, i.e. the fraction of energy available to excite a given ion. 

For the radioactive power model, $L \propto (1 - {\rm e}^{-\tau}) \times M_{\rm Ni}$ is expected, where $\tau$ is the optical depth to the decay gamma rays. In nebular phase, $\tau \ll 1$, which gives $1 - {\rm e}^{-\tau}$ $\sim \tau$, therefore $L \propto \tau \times M_{\rm Ni}$ \citep{maeda03}. At a given phase (in this work, 200 days after $V$-band maximum is reached), 
\begin{equation}
  \tau \propto \frac{M_{\rm ejecta}^2}{E_{\rm K}} \sim M_{\rm ejecta} . 
 \label{eq:tau_Mej}
\end{equation}
Here, we assume that the variation in $M_{\rm ejecta} / E_{\rm K}$ is negligible given that the photospheric velocities from light curve and early spectra modeling are similar among SNe IIb (\citealt{lyman16}, and reference therein). We therefore have the following expression:  
\begin{equation}
   L \propto M_{\rm ejecta} \times M_{\rm Ni} \ .
 \label{eq:Lum_t}
\end{equation} 

\begin{figure*}[!t]
\includegraphics[width = 15cm]{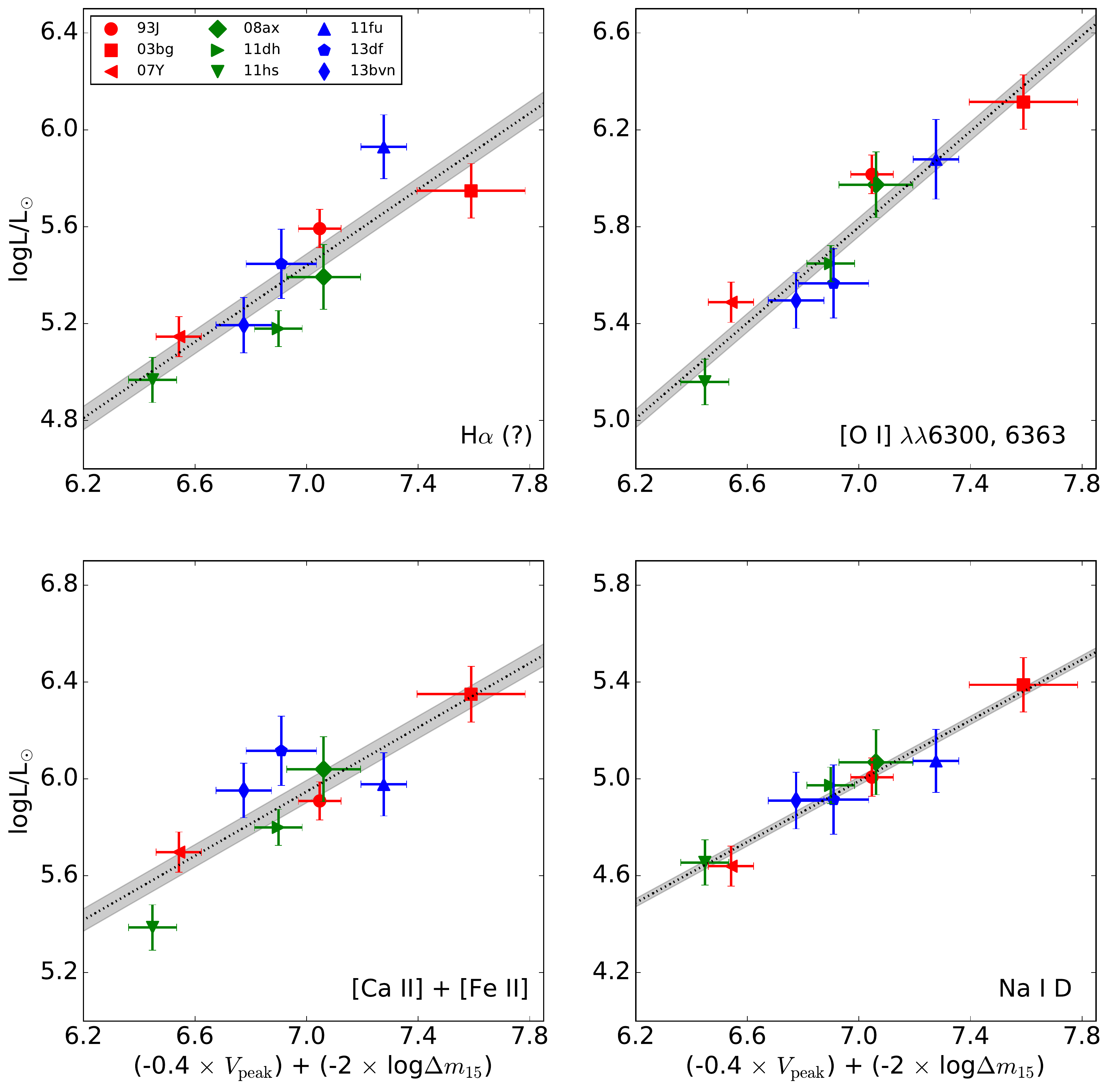}
\centering
\caption{Relative gamma-ray deposition luminosities (see Equation~\ref{eq:L_MNiMej}) versus the luminosities of the nebular lines in logarithmic scale, at $200$ days after the $V$-band maximum. Error in distance is included. In each plot, the dashed line is the best-fit result and shadow region is the standard deviation of the fitting.}
\label{fig:Ni_luma_ejecta}
\vspace{4mm}
\end{figure*}

Figure~\ref{fig:Ni_luma} does not contain information on the possible diversity of the ejecta mass, which would be partly related to the diversity in the light curve decline rates in the bottom panel of Figure~\ref{fig:lc_div} among different objects. To test the effect of different ejecta mass, the line luminosities are plotted against $M_{\rm ejecta} M_{\rm Ni}$ in logarithmic scale in Figure~\ref{fig:Ni_luma_ejecta}. This value in the $x$-axis is estimated by the observables as follows: 
\begin{equation}
  {\rm log}M_{\rm Ni}M_{\rm ejecta} \sim -0.4 \times V_{\rm peak} - 2 \times {\rm log}\Delta m_{15} \ .
 \label{eq:L_MNiMej}
\end{equation}

As shown in Figure~\ref{fig:Ni_luma_ejecta}, the additional (possible) correction for the ejecta mass does not break the correlations, and indeed the scatters seen in the nebular line luminosities decrease. In general, the nebular line luminosities are found to be consistent with the radioactive-power scenario. There is a certain level of the scatter, which possibly comes from a diversity in the masses of the emitting regions of various lines (but not correlated with $M_{\rm Ni}$). Alternatively, the ejecta mass estimated by the early-phase light curve may miss the internal structure affecting the gamma-ray deposition \citep{maeda03}. 

\begin{figure*}[!t]
\includegraphics[width = 19cm]{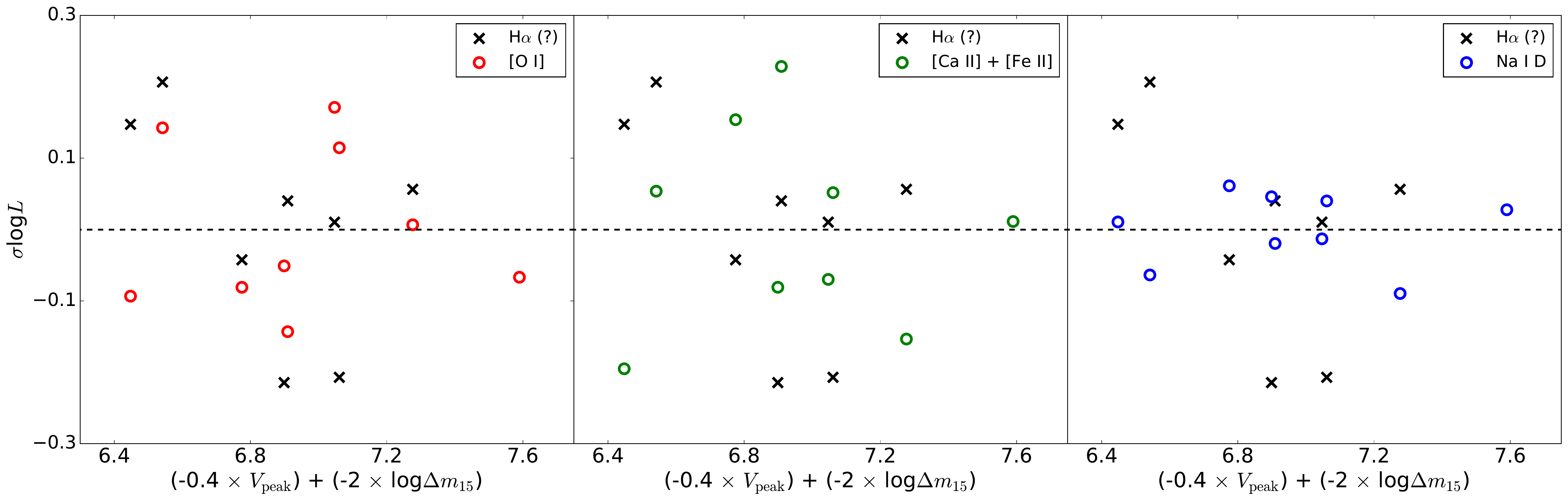}
\centering
\caption{Luminosity scatter level of H$\alpha$-like structure is compared with other emission lines. The H$\alpha$-like structure is labeled by black cross, while other emission lines are labeled by open circles. The H$\alpha$-like structure has approximately same luminosity scatter as [O I] doublet and [Ca II] + [Fe I] complex, but larger than that of Na I D.}
\label{fig:Ni_lumb}
\vspace{4mm}
\end{figure*}

In Figure~\ref{fig:Ni_lumb}, luminosity scatter is compared among different emission lines. The H$\alpha$-like structure gives a similar level of the scatter with the other nebular lines. The envelope mass varies significantly among SNe IIb, from 0.5 $\sim$ 1 $M_{\odot}$ for SN 1993J \citep{shigeyama94} to $\sim$ 0.1 $M_{\odot}$ for SN 2011dh \citep{bersten12}. If the emission would mainly come from the hydrogen envelope through the $\gamma$-ray deposition, we would expect that the line luminosity is roughly proportional to the mass of the emitting region for the following reasons: (1) The ejecta are optically thin to the $\gamma$-rays. Therefore, the deposition rate at a given layer is proportional to its optical depth (i.e., column density) to $\gamma$-rays, which is scaled as $\propto M_{\rm H}/V_{\rm H}^2$ where $M_{\rm H}$ is the H-rich envelope mass and $V_{\rm H}$ is the characteristic velocity of the H-rich envelope (\citealt{kozma92}; \citealt{maeda03}; \citealt{maeda07}). The dispersion in $V_{\rm H}$ can be inferred from the H$\alpha$ absorption velocity around the maximum light. The scatter in the H$\alpha$ absorption velocity is at most within a factor of 1.5 for different SNe IIb \citep{liu16}, which is negligible as compared to the variation in $M_{\rm H}$. The deposited luminosity is therefore roughly proportional to the mass of the hydrogen envelope in this case. (2) The ejecta is optically thin to optical photons. Therefore, the deposited energy is instantaneously converted to optical photons, and the luminosity from a given layer is proportional to its mass. (3) We expect H$\alpha$ is produced through recombination following the non-thermal ionization. In this case, the line luminosity is insensitive to the thermal condition, and roughly scaled by the deposited luminosity, i.e. the mass of the H envelope. Given the above reasons, we expect that the luminosity scatter level would reach to $\sim 0.7 - 1$ dex, much larger than the root mean square (rms) $\sim 0.13$ dex we find in our sample, or larger than the maximum level of the difference between the objects ($\sim 0.4$ dex).  Therefore, we conclude that this emission comes from the inner layer, rather than the hydrogen-rich envelope. 

\citet{prentice17} reach to a similar conclusion by a different approach. They compare the velocities of emission and absorption features, and find that the velocity of this H$\alpha$-like structure is always lower than the H$\alpha$ absorption velocity around the maximum but consistent with the He velocity. They conclude that this H$\alpha$-like structure more likely comes from the He layer, which is consistent with our result.

According to the analyses in this section, we conclude that (1) the power source to the H$\alpha$-like structure is the radioactive decay input, and (2) this is emission from the inner layer (e.g., He layer), not from the H-rich layer. Robust identification of the nature of the emission (i.e., ion) from the current phenomenological approach is not easy, and thus we rely on insight obtained through theoretical investigation. In spectral synthesis simulations \citep{jerk15}, the only other candidate proposed for this feature so far is [N II]. Our finding is in line with this identification.

\subsection{Further analysis of the physical properties: the ratio of [O I] to [Ca II]}
Our previous discussion is model free and mainly based on the observations. However, further insight can be obtained by connecting the observables and the physical parameters. For the $^{56}$Ni mass, its relation to the peak magnitude has been intensively studies and well established \citep[e.g.,][]{lyman16}. The ejecta mass (with a combination of the kinetic energy) can be connected to $\Delta m_{15}$ following the frequently-adopted argument of the diffusion time scale and the shape of the light curve around the maximum phase. However, this relation has not been intensively tested from observational quantities which are independent from the early-phase light curve. 

\begin{figure}[!t]
\includegraphics[width = 9cm]{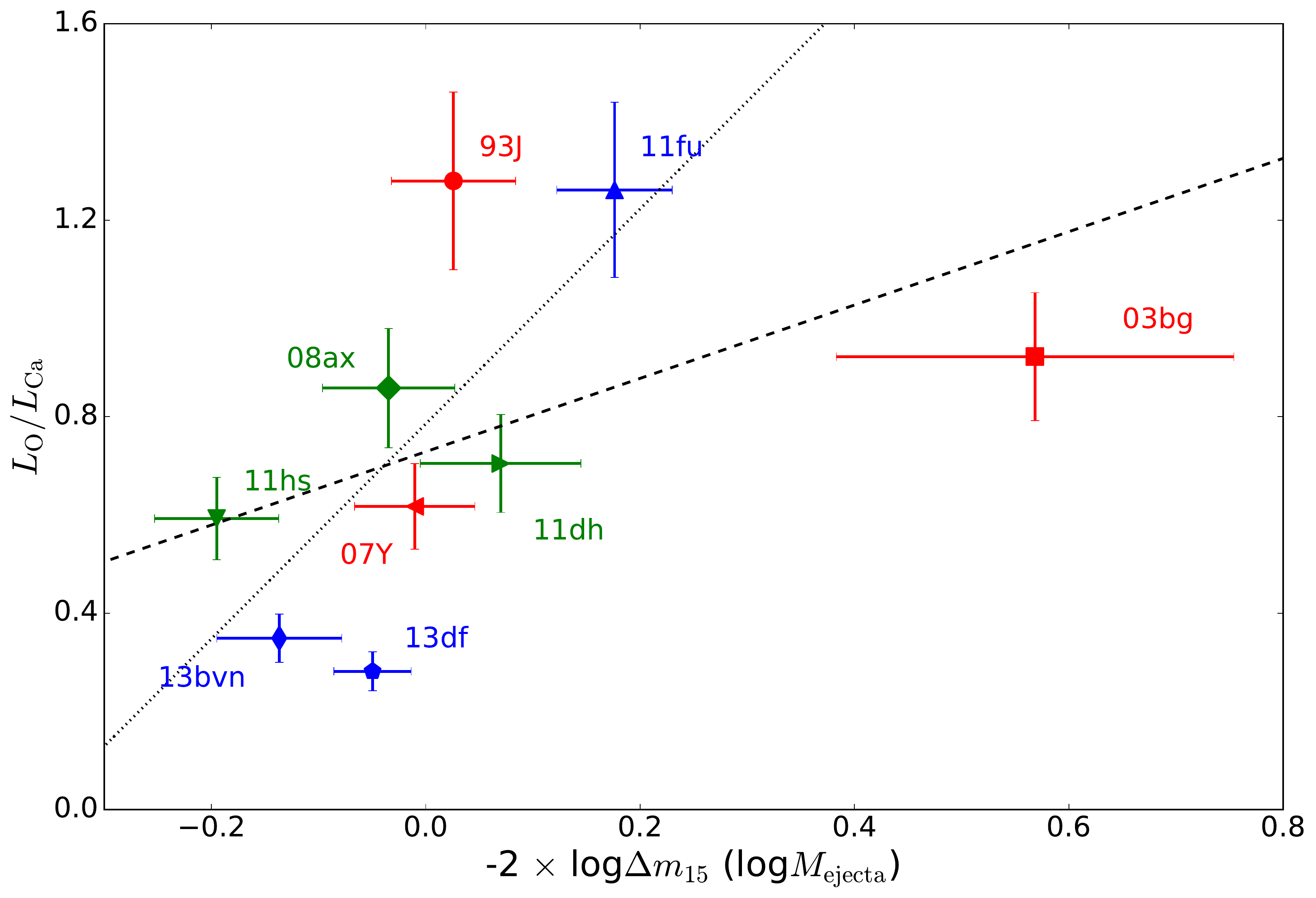}
\centering
\caption{The ejecta mass, estimated from V band light curve, versus the line ratio of [O I] to [Ca II]. The dashed line is the best fit for all the objects, while the dotted line is the fit where SN 2003bg is excluded. Different objects are labeled by the last two digits of the explosion year and letter(s).}
\label{fig:line_ratio}
\vspace{4mm}
\end{figure}

It has been proposed that the ratio of [O I] to [Ca II] (similarly [O I] per energy deposition) can be a tracer of the progenitor mass (\citealt{fransson89}; \citealt{maeda07}; \citealt{elmham11}; \citealt{jerk15}; \citealt{kun15}). Figure~\ref{fig:line_ratio} compares $\Delta m_{15}$ and the line ratio ($L_{\rm O}/L_{\rm Ca}$). The black-dashed line is the linear fit to all these points. A weak correlation (Pearson coefficient $r = 0.47$) can be discerned, and SN 2003bg seems to be an outlier. If we omit SN 2003bg in our fitting, the correlation will become more significant ($r = 0.68$, as shown by the black-dotted line). Objects with a small value of $-2 \times {\rm log} \Delta m_{15}$ (i.e., faster decliners) tend to have a smaller value of $L_{\rm O}/L_{\rm Ca}$. 

The behavior is in line with the idea that a more massive progenitor (with a larger oxygen core) tends to have more massive ejecta. Indeed, we emphasize that this diagnostics of the [O I]/[Ca II] ratio (or [O I] alone) for the progenitor mass has not been tested for a sample of striped envelope SNe, and Figure~\ref{fig:line_ratio} is the first attempt to clarify that the relation exists in the observational data. The scatter is however still present, which might indicate that the ejecta mass is however not a single function determining the efficiency of the gamma-ray deposition in late phases \citep{maeda03}. Still, the present analysis suggests that the ejecta mass is a main parameter to determine the gamma-ray deposition rate, and we can assume that a more massive progenitor produces more massive ejecta. Therefore, we conclude that $\Delta m_{15}$ can be a tracer of the ejecta mass to some extent. 

\section{Discussion}
\citet{cheva10} categorize SNe IIb into two groups: extended SNe IIb (eIIb) and compact SNe IIb (cIIb)\footnote{Note however that this terminology has been revised by several authors. For example, \citet{maeda15} prefer the 'more extended' and 'less extended' progenitors.}. SNe eIIb are expected to have a more massive and extended hydrogen envelope than SNe cIIb and their light curves have two peaks at early phase. \citet{benami15} also reveal the different properties of these two groups. By comparing ultraviolet spectra of SNe IIb in early phase, they find that objects with a double-peak light curve (SN 1993J and SN 2013df) have strong UV excess, while for objects with a single-peak light curve (SN 2001ig and SN 2011dh) such excess is absent (or relatively weak). They attribute this feature to different amounts of CSM around the progenitors, and thus different intensities of the shock-CSM interaction. \citet{maeda15} also find a possible correlation between the extent of the hydrogen envelope of the progenitors and the amount of CSM for SNe IIb. 

These works highlight the difference among the two SN IIb groups. In this section, we will discuss whether these differences (the amount of the hydrogen envelop and different intensities of shock-CSM interaction) would contribute to the emergence of the H$\alpha$-like structure.

\subsection{Flattening of H$\alpha$-like structure in very late phase}
\begin{figure*}[!t]
\includegraphics[width = 15cm]{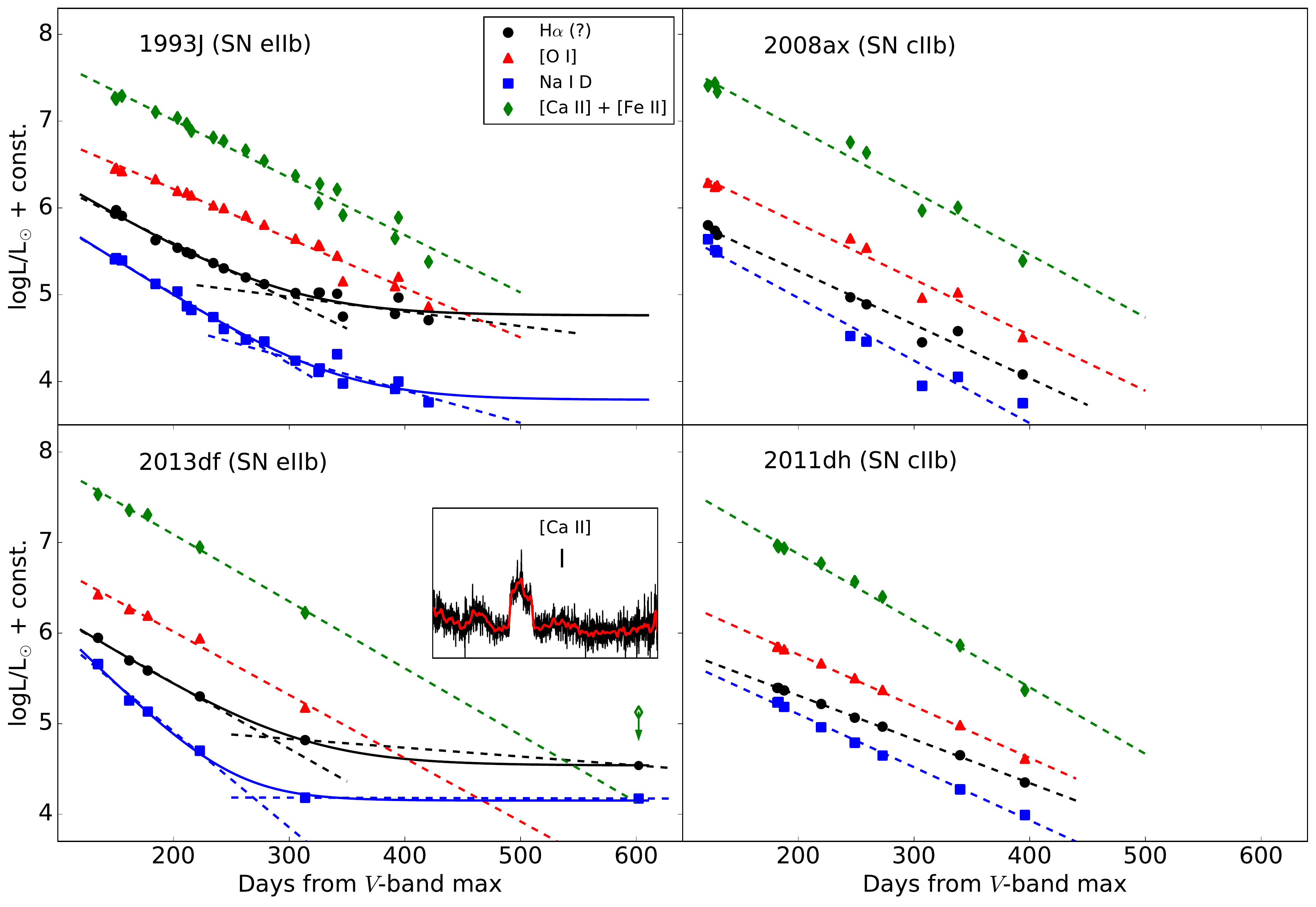}
\centering
\caption{The evolution of the luminosities of nebular lines into the very late stage. Different emission lines are labeled by different colors and markers. Dashed lines are best-fit results (see text for more details). Inner panel shows the spectrum of SN 2013df at $\sim$ 600 days, where the red solid line is its smoothed version. The H$\alpha$-like features and Na I D seen in the hydrogen-rich SNe 1993J and 2013df are flattened at $\> 300$ days after the $V$-band maximum, while [O I] and [Ca II] does not show such a transition. The solid lines are best-fitted results when constant interaction luminosities are included. For the hydrogen-poor SNe 2008ax and 2011dh, the luminosities of all emission lines linearly decline.}
\label{fig:late_flatten}
\vspace{9mm}
\end{figure*}

In previous sections, we conclude that at $\sim$ 200 days, the H$\alpha$-like structure is powered by the radioactive decay of $^{56}$Co, and [N II] is a promising candidate \citep{jerk15}. However, some previous works suggest that this emission line becomes increasingly strong at later epochs for some objects, highlighted by the extended SNe IIb (SNe eIIb) 1993J and 2013df. At later epochs ($> 300$ days), this feature is unambiguously dominated by the shock-CSM interaction (\citealt{matheson00a}; \citealt{weiler07}; \citealt{maeda15}). In this work, the evolution of line luminosity into very late phase ($400$ days after the $V$-band maximum is reached) is also investigated, but only 4 objects (SN 1993J, SN 2008ax, SN 2011dh and SN 2013df) have both photometric data and nebular spectra from early ($<$ 200 days) to late ($>$ 400 days) phases. The result is shown in Figure~\ref{fig:late_flatten}. Note that the logarithm luminosities are shifted by a constant in this figure for illustration purpose, since only the trend in the evolution is important for discussion in this section.

For all the objects, the luminosities of emission lines of interest in this work linearly decline in logarithmic scale before $\sim$ 300 days. However, for the two SNe eIIb, the decline of the luminosities of the H$\alpha$-like structure and Na I D is significantly flattened after $\sim 300$ days. In the case of SN 1993J, data points at $<$ 300 days and $>$ 330 days are fitted separately, and the change in the slope is evident for both emission lines. For SN 2013df, the luminosity of the H$\alpha$-like structure and Na I D at $\sim$ 600 days is obtained by simply integrating the spectrum shown in the inner panel (which is subtracted from \citealt{maeda15}). Here we assume that [O I] is too faint to contribute to the H$\alpha$-like feature significantly at this epoch \citep{maeda15}. A similar flattening takes place at $\sim$ 300 days for both of the SNe eIIb, while such a transition is absent in their relatively compact counterparts up to $\sim$ 400 days. 

Such a flattening is absent in the evolution of the [O I] and [Ca II] luminosity for all SNe IIb, irrespective of the nature of the progenitor. This conclusion may not be evident for SN 2013df, as the noise of the spectrum at $\sim$ 600 days is too large for the luminosity of [Ca II] to  be accurately calculated. However, the luminosity of [Ca II] at $\sim 200$ days is almost an order of magnitude larger than that of the H$\alpha$-like structure (Figure~\ref{fig:line_evo}). Therefore, if such a flattening would have occurred, [Ca II] should have remained sufficiently bright to be easily detected. To estimate the [Ca II] luminosity for SN 2013df at $\sim$ 600 days, we smooth the spectrum by convolving it with Gaussian kernel (red solid line in inner panel of Figure~\ref{fig:late_flatten}), and the luminosity of [Ca II] is calculated by integrating fluxes at few hundreds angstrom around 7300 \AA. Given that the signal-to-noise is too low to distinguish the (possible) emission feature from the background, the estimated [Ca II] luminosity should be regarded as an upper limit. The derived upper limit rejects the possibility of the flattening of the [Ca II] luminosity similar to those of the the H$\alpha$-like feature and Na I D.

Given that SNe eIIb tend to have strong shock-CSM interaction, the behaviors in the luminosities of H$\alpha$-like structure (and Na I D), compared to [O I] and [Ca II], indicate a change of the energy source from the radioactive-power at $\sim$ 200 days to the interaction-power at $\sim$ 400 days. The H$\alpha$-like structure at $\sim 200$ days, which is identified as [N II] in the prior sections, is possibly contaminated by H$\alpha$. Similarly, Na I D may already be contaminated by He I 5876 powered by the shock-CSM interaction. The shock-CSM interaction may thus provide an additional power to these lines, even if this is not a major power source.

To estimate the interaction power at $\sim 200$ days, we assume that the logarithmic luminosity from the radioactive power decays linearly, and the input from shock-CSM interaction is constant. The solid lines are fitted results. By this extrapolation of the shock-CSM interaction power back to $\sim 200$ days, we estimate that the fractions of 15\% (SN 1993J) and 12\% (SN 2013df) in the H$\alpha$-like structure come from the shock-CSM interaction at $\sim 200$ days, and the fractions of 6\% (SN 1993J) and 17\% (SN 2013df) in the `Na I D' are contributed by the shock-powered He I 5876

\subsection{Possible contribution from other sources to the H$\alpha$-like structure}
We now investigate whether the H$\alpha$-like structure is additionally contributed by another mechanism (for example, H$\alpha$ powered by the shock-CSM interaction or by the radioactive decay input). The luminosity scatter of the nebular lines at different epochs are compared for this purpose .

The line luminosities at 5 epochs are compared: 150, 200, 250, 300 and 350 days after the $V$-band maximum. Most of the line luminosities at a given epoch are estimated from a linear fit (Figure~\ref{fig:line_evo}). The analysis at 150 days omits iPTF 13bvn, since the earliest nebular spectra in our sample for this SN was taken at 234 days and the estimation of luminosity at 150 days from linear fit can be very uncertain. SN 2011hs is also excluded in the comparison at 300 days for the same reason. At 350 days, only 5 objects (SNe 1993J, 2008ax, 2011dh, 2013df and iPTF 13bvn) are compared because of a lack of such late-phase spectra for the other objects. Here we note that at $\sim$ 300 days, for SNe 1993J and 2013df, the evolution of the luminosity of H$\alpha$-like structure and Na I D is flattened (see Figure~\ref{fig:late_flatten}), and the linear fit to the line luminosity evolution will underestimate the luminosities. Therefore, for SNe 1993J and 2013df, luminosities of the H$\alpha$-like structure and Na I D are estimated from interpolation.    

\begin{figure*}[!t]
\includegraphics[width = 18cm]{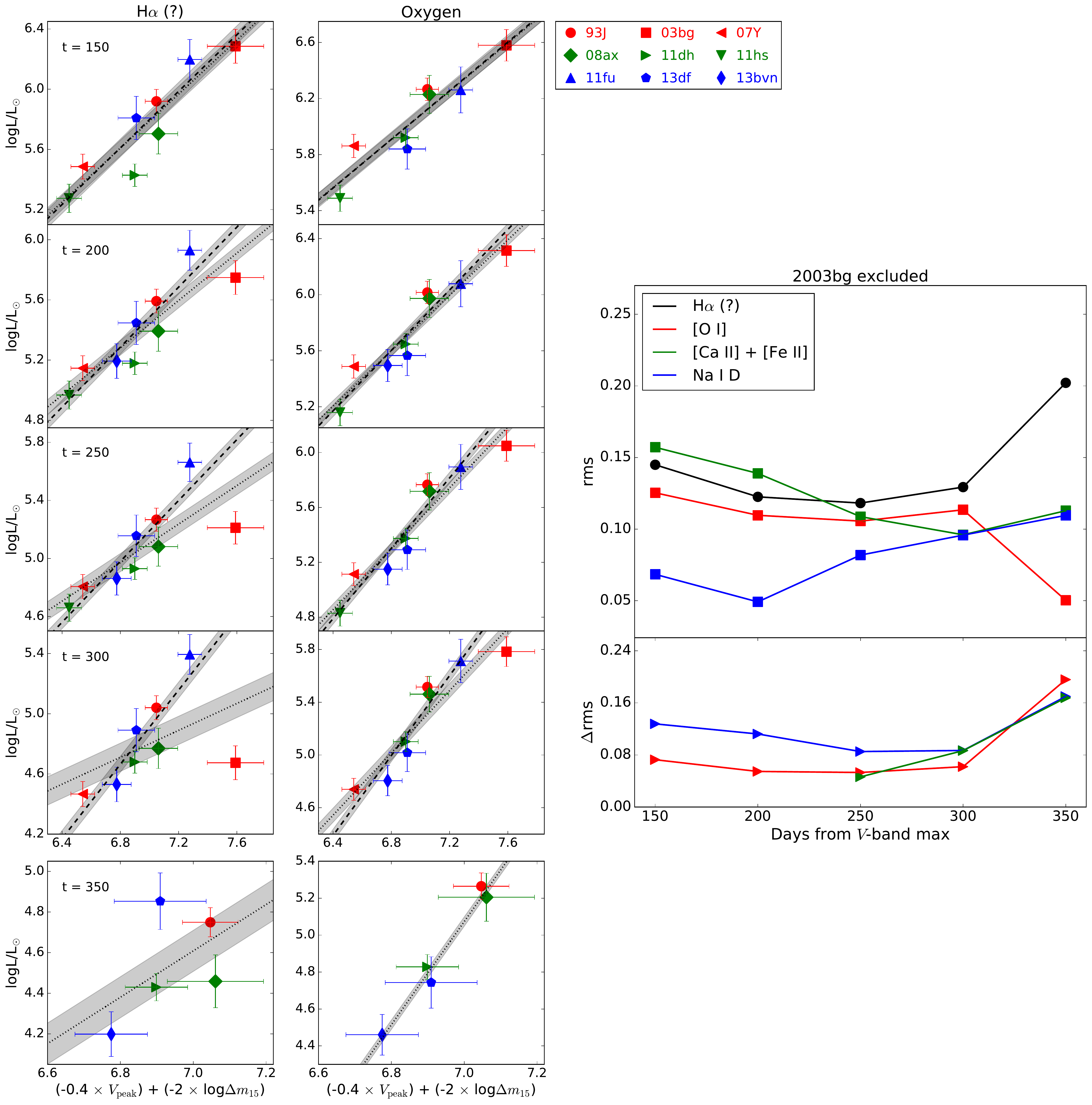}
\centering
\caption{Left panels: deposited energy v.s. line luminosity at different epochs. Only H$\alpha$-like structure and [O I] doublet are illustrated. Dashed lines are fitting results when SN 2003bg is included, while dotted lines are results when SN 2003bg is excluded. Shaded regions are 1$\sigma$ deviations of fitting. Right panel: evolution of rms of different emission lines and the quadratic differences of the rms of the H$\alpha$-like structure and other nebular lines.}
\label{fig:epoch_rms}
\vspace{9mm}
\end{figure*}

Figure~\ref{fig:epoch_rms} (left panels) compares the deposited gamma-ray luminosity equivalent with the line luminosities of the H$\alpha$ structure and [O I] at different epochs. We note that for SN 2003bg, the luminosity of the H$\alpha$-like structure evolves unusually fast, and it seems to be an outlier (also see Figure~\ref{fig:line_evo}). However, detailed discussion on SN 2003bg is beyond the scope of this paper. The dashed lines in the left panels in Figure~\ref{fig:epoch_rms} are the best-fit results when SN 2003bg is excluded, while the dotted lines are the results when SN 2003bg is included. The shaded regions show the 1$\sigma$ deviation of the fitting. The evolution of root mean square (rms) and the quadratic difference between the rms of the H$\alpha$-like and those of other emission lines are also illustrated in the right panel of Figure~\ref{fig:epoch_rms} (for which SN 2003bg is excluded). The Pearson correlation coefficients r are listed in Table~\ref{tab:pearson}

There are three interesting features seen in Figure~\ref{fig:epoch_rms}: (1) At epochs t $<$ 300 days, the luminosities of all the emission lines are correlated with the deposited gamma-ray luminosity. At = 350 days, the luminosities of [O I] and [Ca II] still correlate with the deposited gamma-ray luminosity (r = 0.98 and 0.86 respectively), while for the H$\alpha$-like structure and Na I D, no clear correlation can be discerned anymore (r = 0.51 and -0.15 respectively). (2) At epochs t $<$ 250 days, the H$\alpha$-like structure, [O I] and [Ca II] show a similar level of dispersion. At $t = 350$ days, the luminosity spread of the H$\alpha$-like structure is larger than other emission lines, and the quadratic difference is increasing after 250 days. A similar behavior is seen in Na I D. (3) If SNe with similar deposited gamma-ray luminosities are compared (i.e., SN 1993J v.s. SN 2008ax; SN 2011dh v.s. SN 2013df), the extended SNe IIb (SNe 1993J and 2013df) show more luminous H$\alpha$-like structure than their compact counterparts. 

These features can be interpreted by introducing additional contribution at 200 days from another source, which is linked to the properties of the hydrogen envelope. As the contribution from the shock-CSM interaction is evident after $t \sim 350$ days, this mechanism is a promising candidate. We estimate from the light curve evolution that the fraction of the contribution from the shock-CSM powered H$\alpha$ is $\sim$15\% at 200 days (\S 4.1), which is smaller than the 60-80\% difference seen in the H$\alpha$-like structures between SNe eIIb and SNe cIIb (the point 3 above). Another indication against this interpretation is that there is no gradual increase of the H$\alpha$-like feature up to $t = 300$ days. These features are against the shock-CSM interaction as an additional source of the power at $\sim 200$ days, while a caveat is that this argument is dependent on the time evolution of the shock-CSM interaction power, which is not yet well established. Alternatively, the feature may be contaminated by H$\alpha$ but powered by the radioactive decays. In \S3.1 we concluded that this is not a main contributor, but the variation in the H envelope mass, through different deposition efficiency, may still explain the difference in the flux of the H$\alpha$-like feature between SNe eIIb and cIIb. In any case, our sample is small, and the difference between SNe eIIb and cIIb in the H$\alpha$-like structure at $\sim 200$ days is not statistically significant but only indicative. A larger sample size of SN IIb with high quality nebular data is required to reach to a firm conclusion.

\begin{table}[!b]
\begin{center}
\caption{Pearson correlation coefficients}
\label{tab:pearson}
\begin{tabular}{|l|c|cccc|}
\hline
\multicolumn{2}{|c|}{Lines}&H$\alpha$ (?)&[O I]&[Ca II]&Na I D\\
 \hline
\multirow{2}{*}{150 days}&03bg included&0.92& 0.93 &0.81&0.95\\  
&03bg excluded&0.87&0.88&0.65&0.91\\ \hline
\multirow{2}{*}{200 days}&03bg included&0.88& 0.95&0.86&0.97\\ 
&03bg excluded&0.90&0.93&0.77&0.95\\ \hline
\multirow{2}{*}{250 days}&03bg included&0.78& 0.93&0.92&0.94\\ 
&03bg excluded&0.92&0.95&0.90&0.89\\ \hline
\multirow{2}{*}{300 days}&03bg included&0.47& 0.95&0.90&0.87\\ 
&03bg excluded&0.90&0.94&0.91&0.73\\ \hline
350 days&--&0.51&0.98&0.86&-0.16\\
\hline
\end{tabular}
\end{center}
\end{table}

\section{Conclusions}
In this work, we have analyzed the photometric data and nebular spectra compiled for 7 SNe IIb and 2 SNe Ib. We have investigated the power source of the nebular lines, including the origin of the late-time H$\alpha$-like structure seen in these objects. We have further investigated a possible origin of the diversity among these events from a statistic and model-independent approach. 

In previous works, several scenarios have been proposed as the origin of the H$\alpha$-like structure seen in nebular spectra of SN IIb, at $\sim 200$ days after the maximum brightness. We find a correlation between the luminosity of this emission feature and the mass of $^{56}$Ni produced in the explosion, which is not expected for the shock-CSM interaction scenario. This points to the radioactive decay of $^{56}$Co as a predominant power source of this feature. Further, our analysis clarifies that a level of the scatter in the luminosities of this feature is similar to those of the other metal lines. This is not consistent with the idea that the feature is H$\alpha$ powered by the radioactive decay, as in this case the diversity in the mass of the hydrogen envelope among SNe IIb would create a larger scatter in the luminosities of this H$\alpha$-like feature than other lines. This is further supported by the mass of the hydrogen envelope itself generally inferred for SNe IIb, which is not enough to produce such a luminous emission line \citep{patat95}. We therefore conclude that this line is mainly emitted from the inner layer, and powered by radioactive decay chain. Since [N II] in the He layer is the only candidate from simulations so far, we attribute the origin of this emission as [N II]. While we are not able to robustly exclude other possibilities from our phenomenological and observational approach, this identification provides a picture consistent with the observational constraints we have investigated in this paper.

Our conclusion on the origin of this H$\alpha$-like emission, as dominated by [N II] powered by the radioactive decay of $^{56}$Co, is in line with the nebular spectral synthesis models by \citet{jerk15}. In addition to our main analyses presented in this paper, We have further investigated if the variation expected for the masses of N/He layer would be seen as a scatter in the sample of SNe IIb (plus 2 SNe Ib). We do not see such a variation, suggesting either that the progenitor mass range is relatively small for SNe IIb or that the expected variation is absorbed in the dependence on the progenitor mass, or both. Identifying the origin of this feature as [N II] thus has an interesting avenue for further investigation, i.e., a possible difference in the progenitor mass range for SNe IIb, Ib, and Ic, which we will present in a forthcoming paper (Fang et al., in prep.). The H$\alpha$-like structure is also presented in nebular spectra of some SNe Ib (e.g., SN 2007Y and iPTF 13bvn analyzed in this work, and SN 2007C, see~\citealt{tauben09}). If it is [N II] powered by the radioactive decay, whether such an emission is present can be used as an indicator of the level of the He layer stripping. 

We also find a possible systematic difference in the strengths of the H$\alpha$ structure between the extended SNe IIb (SNe 1993J and 2013df) and the compact ones (SNe 2008ax and 2011dh), already at $\sim 200$ days before the clear shock-CSM signature is observed for the former ($t > 300$ days). The luminosity evolution in very late stage is also compared among different objects. The logarithmic luminosities of the H$\alpha$-like structure and [Ca II] linearly decline before $\sim 250$ days. However, for SNe IIb with an extended envelope, i.e. SNe 1993J and 2013df, a transition takes place at $\sim$ 300 day where the evolution of luminosity of the H$\alpha$-like structure is significantly flattened. In contrast, such a flattening is absent for SNe IIb with a less extended envelope up to $\sim$ 400 days. The luminosities of [O I] and [Ca II] continue to drop for all SNe IIb for which the analysis is possible. The flattening of the evolution of H$\alpha$-like structure luminosity is interpreted as a result of a transition of energy source from radioactivity to shock-CSM interaction. However, the expected level of the shock-CSM contribution is not consistent with the difference between SNe eIIb and cIIb at 200 days, if we assume luminosity from the shock-CSM interaction is constant. Alternatively, the difference may simply reflect the variation of the masses in the H envelope. In any case, this work extends the intrinsic difference among SN IIb, and concludes that the two types of SNe IIb behave differently in nebular phase, which is a topic to be investigated in a forthcoming work (Fang et al. in prep). 

As an additional analysis, we further investigate a relation between the line ratio of [O I]/[Ca II] and the post-maximum light curve decline rate. The correlation exists, and this finding suggests that the line ratio [O I]/[Ca II] can be an indicator of the progenitor mass and ejecta mass. A systematic study of this line ratio to a sample of SNe IIb/Ib/Ic will be presented in a forthcoming paper (Fang et al. in prep). 

We note that our result is limited by the relatively small sample size. Our future work aims at enlarging the sample of both nebular spectra and photometric data of SNe IIb, and further extending to SNe Ib and Ic. The analyses we present in this paper can form a solid basis to apply to a larger sample of SNe IIb/Ib/Ic. 

\acknowledgements
QF acknowledges support by MEXT scholarship awarded by Ministry of Education, Culture, Sports, Science and Technology, Japan. KM acknowledges support by Japan Society for the Promotion of Science (JSPS) KAKENHI Grant 17H02864 and 18H04585. We thank Zhuo Li, Jinyi Shangguan and Herczeg Gregory for stimulating discussion and helpful comments. We thank Antonio Morales-Garoffolo for kindly providing us with the spectra of SN 2013df. We thank the anonymous referee for many constructive comments to improve the analysis. We thank the Weizmann interactive supernova data repository (WISeREP) for access to supernova data. 


\end{CJK*}

\end{document}